\documentclass[twocolumn,11pt]{article}

\usepackage[utf8]{inputenc}
\usepackage[margin=1in]{geometry}
\usepackage{setspace}
\usepackage{amsmath, amssymb, amsfonts}
\usepackage{algorithm}
\usepackage{algorithmic}

\usepackage{graphicx}
\usepackage{booktabs}
\usepackage{multirow}
\usepackage{tabularx}
\usepackage{hhline}
\usepackage{pifont}
\usepackage{subcaption}
\usepackage{caption}
\usepackage{pgfplots}
\pgfplotsset{compat=1.18}

\usepackage{textcomp}
\usepackage[normalem]{ulem}        % for \sout
\usepackage{soul}                  % for \hl
\usepackage{xcolor}

\usepackage[colorlinks=true, linkcolor=blue, citecolor=blue, urlcolor=blue]{hyperref}

\usepackage{cite}

\begin{document}
\title{MD-ViSCo: A Unified Model for Multi-Directional Vital Sign Waveform Conversion}

\author{
    Franck Meyer$^{1,*}$, 
    Kyunghoon Hur$^{1,*}$, 
    Edward Choi$^{1}$ \\
    \\
    $^1$Kim Jaechul Graduate School of AI, KAIST, South Korea \\
    \texttt{\{franck.meyer, pacesun, edwardchoi\}@kaist.ac.kr}
}
\date{\today}
\maketitle
\renewcommand{\thefootnote}{\fnsymbol{footnote}}
\footnotetext[1]{* Equal contribution (co-first authors).}
\renewcommand{\thefootnote}{\arabic{footnote}}

\begin{abstract} Despite the remarkable progress of deep-learning methods generating a target vital sign waveform from a source vital sign waveform, most existing models are designed exclusively for a specific source-to-target pair. This requires distinct model architectures, optimization procedures, and pre-processing pipelines, resulting in multiple models that hinder usability in clinical settings.
To address this limitation, we propose the \textbf{Multi-Directional Vital-Sign Converter (MD-ViSCo)}\footnote{Our code implementation is available on Github - https://github.com/fr-meyer/MD-ViSCo}, a unified framework capable of generating any target waveform such as electrocardiogram (ECG), photoplethysmogram (PPG), or arterial blood pressure (ABP) from any single input waveform with a single model.
MD-ViSCo employs a shallow 1-Dimensional U-Net integrated with a Swin Transformer that leverages Adaptive Instance Normalization (AdaIN) to capture distinct waveform styles. 
To evaluate the efficacy of MD-ViSCo, we conduct multi-directional waveform generation on two publicly available datasets. Our framework surpasses state-of-the-art baselines (NabNet \& PPG2ABP) on average across all waveform types, lowering Mean absolute error (MAE) by \textbf{8.8\%} and improving Pearson correlation (PC) by \textbf{4.9\%} over two datasets.
In addition, the generated ABP waveforms satisfy the Association for the Advancement of Medical Instrumentation (AAMI) criterion and achieve Grade B on the British Hypertension Society (BHS) standard, outperforming all baselines.
By eliminating the need for developing a distinct model for each task, we believe that this work offers a unified framework that can deal with any kind of vital sign waveforms with a single model in healthcare monitoring.
\end{abstract}


\section{Introduction} \label{sec:introduction}
\begin{figure}[]
\centering
\centerline{\includegraphics[width=0.45\textwidth, height=5.5cm]{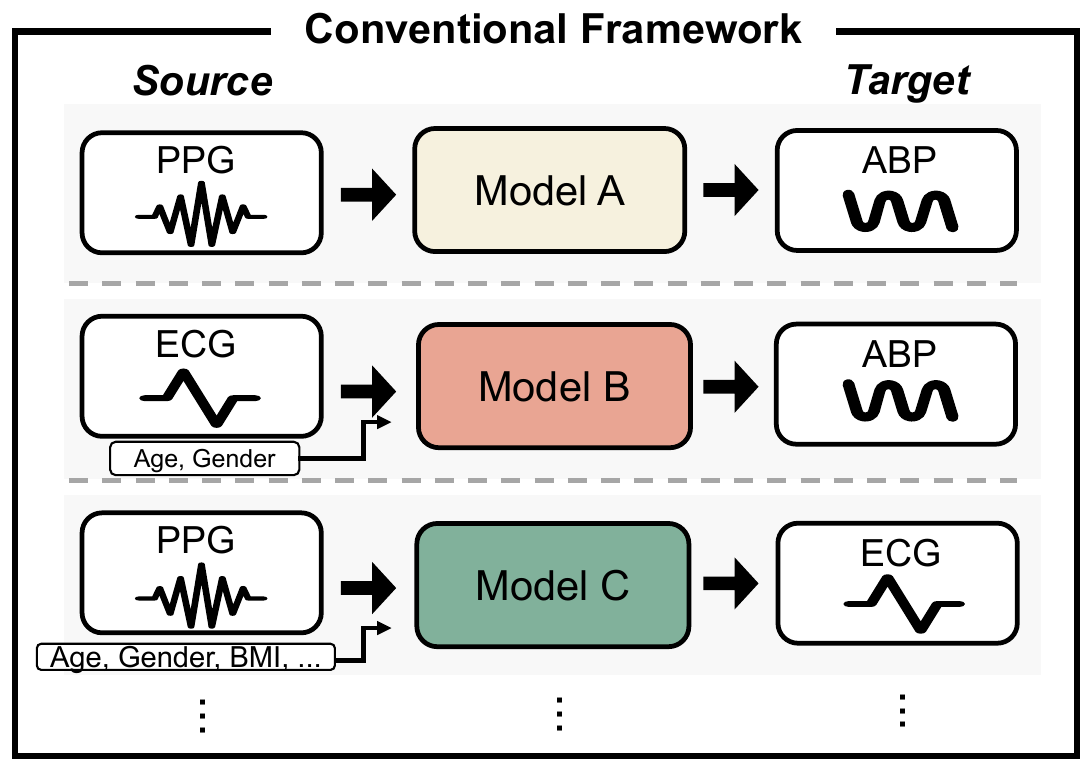}}
\caption{For the conventional waveform conversion framework, a distinct model must be trained for each source‑to‑target direction, each with its own architecture, objective function, and subsets of selected demographic features.}
\label{fig:fig1}
\vskip -5pt
\end{figure}

Continuous and accurate monitoring of various vital signs such as electrocardiograms (ECG), photoplethysmograms (PPG), and arterial blood pressure (ABP) waveforms is critical for tracking patients' status, which in turn facilitates timely and effective medical intervention~\cite{bib46,bib48}.
However, unlike Intensive Care Unit (ICU), the use of comprehensive monitoring systems in general wards or emergency departments is rarely feasible due to limited clinical equipment and personnel, even though continuous monitoring of vital signs is often needed~\cite{bib50}.
For that, the waveform conversion method mitigates this gap by computationally generating missing signals, reducing the reliance on specific measurement~\cite{bib54,bib21}.
One promising use case is the conversion of non-invasive into invasive waveforms (e.g., PPG$\rightarrow$ABP), since traditional methods such as arterial line insertion impose constraints---including the need for a sterile environment, trained medical personnel, and constant supervision---and carry risks such as infection or bleeding~\cite{bib13,bib56}.

\begin{table*}[ht]
\renewcommand{\arraystretch}{1.3}
\centering
\resizebox{0.96\textwidth}{!}{%
\begin{tabular}{llccccccccccccl} % one fewer column 'c'
 &   &
  \rotatebox{70}{\shortstack{\hyperref[bib2]{Athaya et al.}\\(2021)}} &
  \rotatebox{70}{\shortstack{\hyperref[bib14]{Aguirre et al.}\\(2021)}} &
  \rotatebox{70}{\shortstack{\hyperref[bib19]{Vo et al.}\\(2021)}} &
  \rotatebox{70}{\shortstack{\hyperref[bib15]{Dasgupta et al.}\\(2021)}} &
  \rotatebox{70}{\shortstack{\hyperref[bib4]{Ibtehaz et al.}\\(2022)}} &
  \rotatebox{70}{\shortstack{\hyperref[bib6]{Mahmud et al.}\\(2023)}} &
  \rotatebox{70}{\shortstack{\hyperref[bib16]{Tang et al.}\\(2023)}} &
  \rotatebox{70}{\shortstack{\hyperref[bib7]{Zhang et al.}\\(2023)}} &
  \rotatebox{70}{\shortstack{\hyperref[bib36]{Pan et al.}\\(2024)}} &
  \rotatebox{70}{\shortstack{MD-ViSC\\(Ours)}} &
   \\ \cline{1-12}
\multirow{3}{*}{\rotatebox{90}{\bf{Source}}} & PPG  &
  \checkmark &
  \checkmark &
  \checkmark &
  \checkmark &
  \checkmark &
  \checkmark &
  \checkmark &
  \checkmark &
  \checkmark &
  \checkmark \\
 & ECG &
  \textcolor[gray]{0.95}{\ding{55}} &
  \textcolor[gray]{0.95}{\ding{55}} &
  \textcolor[gray]{0.95}{\ding{55}} &
  \textcolor[gray]{0.95}{\ding{55}} &
  \textcolor[gray]{0.95}{\ding{55}} &
  \checkmark &
  \textcolor[gray]{0.95}{\ding{55}} &
  \checkmark &
  \checkmark &
  \checkmark \\
 & ABP &
  \textcolor[gray]{0.95}{\ding{55}} &
  \textcolor[gray]{0.95}{\ding{55}} &
  \textcolor[gray]{0.95}{\ding{55}} &
  \textcolor[gray]{0.95}{\ding{55}} &
  \textcolor[gray]{0.95}{\ding{55}} &
  \textcolor[gray]{0.95}{\ding{55}} &
  \textcolor[gray]{0.95}{\ding{55}} &
  \textcolor[gray]{0.95}{\ding{55}} &
  \textcolor[gray]{0.95}{\ding{55}} &
  \checkmark \\
\cline{1-12}
\multirow{3}{*}{\rotatebox{90}{\bf{Target}}} & PPG &
  \textcolor[gray]{0.95}{\ding{55}} &
  \textcolor[gray]{0.95}{\ding{55}} &
  \textcolor[gray]{0.95}{\ding{55}} &
  \textcolor[gray]{0.95}{\ding{55}} &
  \textcolor[gray]{0.95}{\ding{55}} &
  \textcolor[gray]{0.95}{\ding{55}} &
  \textcolor[gray]{0.95}{\ding{55}} &
  \textcolor[gray]{0.95}{\ding{55}} &
  \textcolor[gray]{0.95}{\ding{55}} &
  \checkmark \\
 & ECG &
  \textcolor[gray]{0.95}{\ding{55}} &
  \textcolor[gray]{0.95}{\ding{55}} &
  \checkmark &
  \checkmark &
  \textcolor[gray]{0.95}{\ding{55}} &
  \textcolor[gray]{0.95}{\ding{55}} &
  \checkmark &
  \textcolor[gray]{0.95}{\ding{55}} &
  \textcolor[gray]{0.95}{\ding{55}} &
  \checkmark \\
 & ABP &
  \checkmark &
  \checkmark &
  \textcolor[gray]{0.95}{\ding{55}} &
  \textcolor[gray]{0.95}{\ding{55}} &
  \checkmark &
  \checkmark &
  \textcolor[gray]{0.95}{\ding{55}} &
  \checkmark &
  \checkmark &
  \checkmark \\
\cline{1-12}
\multirow{3}{*}{\rotatebox{90}{\bf{Dataset}}} &
  MIMIC-II &
  \textcolor[gray]{0.95}{\ding{55}} &
  \textcolor[gray]{0.95}{\ding{55}} &
  \checkmark &
  \textcolor[gray]{0.95}{\ding{55}} &
  \checkmark &
  \checkmark &
  \checkmark &
  \checkmark &
  \checkmark &
  \checkmark \\
 &
  MIMIC-III &
  \checkmark &
  \checkmark &
  \textcolor[gray]{0.95}{\ding{55}} &
  \checkmark &
  \textcolor[gray]{0.95}{\ding{55}} &
  \textcolor[gray]{0.95}{\ding{55}} &
  \textcolor[gray]{0.95}{\ding{55}} &
  \textcolor[gray]{0.95}{\ding{55}} &
  \textcolor[gray]{0.95}{\ding{55}} &
  \checkmark \\
 &
  VitalDB &
  \textcolor[gray]{0.95}{\ding{55}} &
  \textcolor[gray]{0.95}{\ding{55}} &
  \textcolor[gray]{0.95}{\ding{55}} &
  \textcolor[gray]{0.95}{\ding{55}} &
  \textcolor[gray]{0.95}{\ding{55}} &
  \textcolor[gray]{0.95}{\ding{55}} &
  \textcolor[gray]{0.95}{\ding{55}} &
  \textcolor[gray]{0.95}{\ding{55}} &
  \checkmark &
  \checkmark \\
\cline{1-12}
\multirow{3}{*}{\rotatebox{90}{\bf{Metrics}}} &
  Similarity &
  \checkmark &
  \checkmark &
  \checkmark &
  \checkmark &
  \checkmark &
  \checkmark &
  \checkmark &
  \checkmark &
  \checkmark &
  \checkmark \\
 &
  AAMI &
  \checkmark &
  \textcolor[gray]{0.95}{\ding{55}} &
  \textcolor[gray]{0.95}{\ding{55}} &
  \textcolor[gray]{0.95}{\ding{55}} &
  \checkmark &
  \checkmark &
  \textcolor[gray]{0.95}{\ding{55}} &
  \textcolor[gray]{0.95}{\ding{55}} &
  \checkmark &
  \checkmark \\
 &
  BHS &
  \checkmark &
  \checkmark &
  \textcolor[gray]{0.95}{\ding{55}} &
  \textcolor[gray]{0.95}{\ding{55}} &
  \checkmark &
  \checkmark &
  \textcolor[gray]{0.95}{\ding{55}} &
  \textcolor[gray]{0.95}{\ding{55}} &
  \checkmark &
  \checkmark \\
\cline{1-12}
\multirow{2}{*}{\rotatebox{90}{\bf{Modality}}} &
 \rule{0pt}{1.3em}Waveform & 
 \rule{0pt}{1.3em}\checkmark &
 \rule{0pt}{1.3em}\checkmark &
 \rule{0pt}{1.3em}\checkmark &
 \rule{0pt}{1.3em}\checkmark &
 \rule{0pt}{1.3em}\checkmark &
 \rule{0pt}{1.3em}\checkmark &
 \rule{0pt}{1.3em}\checkmark &
 \rule{0pt}{1.3em}\checkmark &
 \rule{0pt}{1.3em}\checkmark &
 \rule{0pt}{1.3em}\checkmark \\
 &
 \rule{0pt}{1.6em}Patient info. &
 \rule{0pt}{1.6em}\textcolor[gray]{0.95}{\ding{55}} &
 \rule{0pt}{1.6em}\checkmark &
 \rule{0pt}{1.6em}\textcolor[gray]{0.95}{\ding{55}} &
 \rule{0pt}{1.6em}\textcolor[gray]{0.95}{\ding{55}} &
 \rule{0pt}{1.6em}\textcolor[gray]{0.95}{\ding{55}} &
 \rule{0pt}{1.6em}\textcolor[gray]{0.95}{\ding{55}} &
 \rule{0pt}{1.6em}\textcolor[gray]{0.95}{\ding{55}} &
 \rule{0pt}{1.6em}\textcolor[gray]{0.95}{\ding{55}} &
 \rule{0pt}{1.6em}\textcolor[gray]{0.95}{\ding{55}} &
 \rule{0pt}{1.6em}\checkmark \\
\noalign{\vskip 0.5em} \cline{1-12}
\multicolumn{2}{l}{Code available} &
  \textcolor[gray]{0.95}{\ding{55}} &
  \checkmark &
  \checkmark &
  * &
  \checkmark &
  * &
  \textcolor[gray]{0.95}{\ding{55}} &
  \textcolor[gray]{0.95}{\ding{55}} &
  * &
  \checkmark \\
\cline{1-12}
\multicolumn{2}{l}{Extensible\textsuperscript{$\dagger$}} &
  \textcolor[gray]{0.95}{\ding{55}} &
  \textcolor[gray]{0.95}{\ding{55}} &
  \textcolor[gray]{0.95}{\ding{55}} &
  \textcolor[gray]{0.95}{\ding{55}} &
  \textcolor[gray]{0.95}{\ding{55}} &
  \textcolor[gray]{0.95}{\ding{55}} &
  \textcolor[gray]{0.95}{\ding{55}} &
  \textcolor[gray]{0.95}{\ding{55}} &
  \textcolor[gray]{0.95}{\ding{55}} &
  \checkmark \\

\end{tabular}
}
\vspace{3pt}

\noindent\hspace{3em}\parbox{\linewidth}{%
\footnotesize For datasets, only those that are publicly available and more than 1,000 samples are included. \\
\footnotesize *: Github repository is presented but does not provide code for reproducing results. \\
\footnotesize $^\dagger$ : Provided interface and clear guidelines to easily add interoperable modules and extend waveform type with minimal refactoring.
}
\caption{Comparison of physiological waveform conversion studies, ordered by publication date.  
MD-ViSCo (Ours) supports all three types of waveforms as both a source and a target waveform. 
Similarity metrics refer to Mean Standard Error(MSE), Mean Absolute Error(MAE), while AAMI and BHS indicate medical application standards.  
Modality indicates whether the model incorporates additional input beyond waveforms; to the best of our knowledge, no prior work uses auxiliary modality other than patient information.  
MD-ViSCo is publicly available and allows the addition of modules and waveform types.}  
\label{tab:comparison_works}
\vskip -5pt
\end{table*}

Recent advances in artificial intelligence(AI) have facilitated these waveform conversion methods~\cite{bib37, bib53, bib2}. 
However, as illustrated in Figure~\ref{fig:fig1}, most current AI models are developed for uni-directional conversion, where each model is designed to transform a specific source into a target waveform (e.g., PPG$\rightarrow$ABP or PPG$\rightarrow$ECG).
This approach requires separate models for each source-to-target conversion, which induces inefficiency, since multiple models with distinct architectures must be trained, optimized, and maintained.
Moreover, as the number of required conversion tasks grows, so does the engineering overhead.
The result is a fragmented landscape of single‑objective models that are neither scalable nor share model parameters across different tasks, raising the clear need for a unified model that can handle any source-to-target waveform conversion~\cite{bib52}.

Meanwhile, recent studies aim to enhance performance by leveraging multiple waveforms simultaneously (e.g. ECG+PPG$\rightarrow$ABP)~\cite{bib7,bib36}.
Although the use of multiple waveforms is promising, these approaches assume that multiple waveform modalities should be available at the same time, an assumption rarely encountered in general wards or pre‑hospital care.
Consequently, there is a need for models that can learn inter‑waveform relationships during training, while facilitating waveform conversion from a single input waveform at the inference stage.

To overcome these challenges, we propose the Multi-Directional Vital-Sign Converter (MD-ViSCo) that uses any single source waveform as input to generate any type of target waveform within a unified model.
Although the model operates in a setting where each sample contains a single waveform type, it can implicitly learn relationships among different waveform types by being trained on a dataset that pools multiple waveform types.
First, \emph{multi-directional approximation model} is designed to generate normalized vital sign waveforms from any single input waveform.
This model enables scalable and efficient conversion across diverse waveform types without requiring the training and management of separate models for each specific conversion direction.
Second, \emph{refinement model} further refines these approximations into real-unit valued units by incorporating patient demographic information with textual embeddings.

\begin{figure*}[ht] 
\centering
\centerline{\includegraphics[width=0.81\textwidth, height=7.7cm]{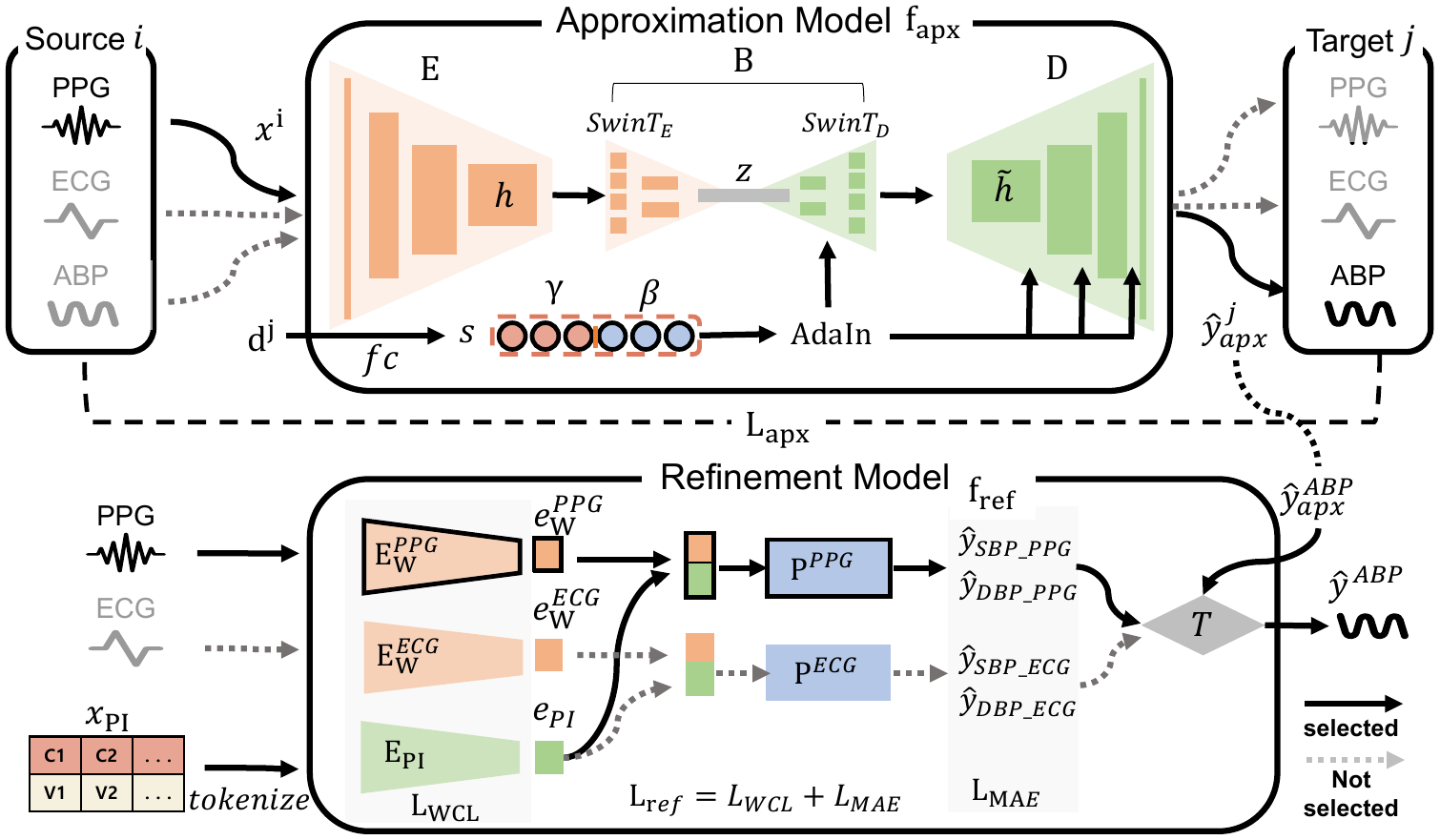}}
\caption{Overview of MD-ViSCo.
The approximation model \(f_{\text{apx}}\) receives a source waveform \(x^{(i)}\) and a target domain indicator \(d^{(j)}\) to generate an normalized target waveform \(\hat{y}_{apx}^{(j)}\).
For arterial blood pressure (ABP) waveform generation, the refinement model \(f_{\text{ref}}\) further processes \(\hat{y}_{apx}^{APX}\) along with patient information \(x_{\text{PI}}\) by predicting systolic and diastolic blood pressure values \((\hat{y}_{\text{SBP}}, \hat{y}_{\text{DBP}})\).
The final ABP waveform \(\hat{y}^{(j)}\) in real-valued mmHg units is obtained via a linear transformation \(T\). During training, all components including dotted arrows are used, whereas during inference, only the solid arrow is active.}
\label{fig:fig2}
\vskip -5pt
\end{figure*}

In this study, we evaluate MD-ViSCo against state-of-the-art (SOTA) baseline models originally developed for specific waveform conversion tasks through two publicly available datasets (UCI~\cite{bib35} and PulseDB~\cite{bib24}) using three evaluation settings; (i) Waveform conversion based on similarity metric, (ii) Extracting features from generated waveform based on its morphology, (iii) Compliance with two key blood pressure (BP) measurement standards for ABP: AAMI (Association for the Advancement of Medical Instrumentation) and BHS (British Hypertension Society).

Our key contributions are summarized as follows:
\begin{itemize}
\item We present a unified vital sign waveform conversion framework that processes any waveform type conversion and refines the generated waveform by scaling it to real units using demographic patient information.
\item Compared to SOTA uni-directional models, our proposed framework achieves superior or comparable performance in waveform conversion (in Figure~\ref{fig:fig3}) and physiological feature fidelity (in Figure~\ref{fig:feature}).
\item We demonstrate its feasibility by showing that it outperforms baseline models according to the AAMI (in Table~\ref{tab:aami_results}) and BHS(in Figure~\ref{fig:fig4}) medical device standards, highlighting its clinical applicability in generating ABP waveforms in millimeters of mercury (mmHg).
\end{itemize}

\section{Related Works} \label{sec:related_works}
\subsection{Deep Learning method for vital sign waveform conversion}
Deep learning techniques have been widely adopted for vital sign waveform conversion.
In particular, methods such as PPG2ABP~\cite{bib4} and NabNet~\cite{bib6} demonstrate how non-invasive waveforms (PPG or ECG) can be transformed into an invasive ABP waveform via a two-stage cascade model, thus enabling non-invasive monitoring as an alternative in situations where invasive methods are required but difficult to implement.
For non-invasive to non-invasive conversions, many studies focus on translating PPG waveforms into ECG waveforms that provide richer cardiac clinical information but are harder to capture than PPG~\cite{bib18,bib22}.
In addition to these representative conversion models, previous studies have employed diverse methodologies, such as transfer learning~\cite{bib9}, frequency domain transforms~\cite{bib12}, adversarial training (GANs)~\cite{bib19, bib15}, and patient-specific trained models~\cite{bib16, bib20}, for waveform conversion tasks.
However, all of these approaches focus on a single target conversion, limiting their scope to a specific task or evaluation conditions (e.g., dataset, metrics, modulus), as described in Table~\ref{tab:comparison_works}.
This leads to disparities in evaluation criteria, model architectures, and maintenance workflows (e.g., updating patient-specific models for unseen patients), making it difficult to compare methods fairly or generalize across settings.

\subsection{Incorporating additional modalities: demographics and multiple waveform modalities}
Apart from methods relying solely on a single waveform as input, recent studies have investigated utilizing multiple data sources to improve waveform conversion performance.
One line of work explores the integration of patient demographic information as an additional modality~\cite{bib14, bib8}.
In practice, since demographic features vary across datasets (e.g., age, sex, weight), model development must be tailored for both preprocessing and architecture.
Another approach adopts multiple waveform modalities, feeding them into the model to capture cross-waveform dependencies via early/late fusion or attention mechanisms~\cite{bib6, bib15}.
However, model performance degrades when one waveform type is missing, which is a scenario common outside the ICU.

\section{Methodology} \label{sec:method}
\subsection{Datasets}
We evaluate our proposed model using two datasets: PulseDB~\cite{bib24} and UCI dataset~\cite{bib35}.
Both datasets contain synchronized ECG, PPG, and ABP waveforms, all sampled at 125Hz, enabling waveform conversion studies.
ECG and PPG are provided in locally min-max normalized form (0–1 range, sample-wise), whereas ABP is retained in real-valued units.
Details on dataset pre-processing are presented in the Appendix~\ref{appendix:dataset}.

\textbf{PulseDB~\cite{bib24}} is constructed by merging VitalDB~\cite{bib38} and the MIMIC-III waveform dataset~\cite{bib39}.
It comprises 5,245,454 segments of 10-second ECG, PPG, and ABP recordings collected from 5,361 patients.
Pre-processing includes signal synchronization, noise removal, zero-centering, and zero padding to ensure a consistent length of 1,280 data points.
It also includes patient demographic information, although the available features differ between VitalDB and MIMIC-III: age and sex in VitalDB, and additional attributes such as height, weight, and BMI in MIMIC-III.

\textbf{UCI Dataset~\cite{bib35}} derived from MIMIC-II Waveform database, contains 12,000 ECG, PPG, and ABP recordings.
Following the original NabNet pre-processing pipeline, the dataset was filtered, synchronized, baseline corrected, and normalized.
After pre-processing, it comprises 191,186 segments with 8.192-seconds(1,024 data points) each.
UCI dataset does not provide any demographic information.

\subsection{Waveform conversion process}
In this section, we describe the waveform conversion process and introduce the notation used throughout this paper.
Each patient record contains a set of single-channel vital sign waveforms of different types, such as ECG, PPG, and ABP, which are recorded simultaneously. 
For waveform conversion between these vital sign types, we define the sets of source and target waveforms as follows:
Let \(\mathcal{X} = \{x^{(i)} \mid i \in \mathcal{T}\}\) and \(\mathcal{Y} = \{y^{(j)} \mid j \in \mathcal{T}\}\), where each \(x^{(i)}, y^{(j)} \in \mathbb{R}^{1 \times L}\) represents a source and target waveform of types \(i\) and \(j\), respectively, and \(\mathcal{T} = \{\text{ECG},\ \text{PPG},\ \text{ABP}\}\) denotes the set of waveform types. 
All waveforms have a same sequence length $L$, defined by the sampling rate and segment duration: \(L = \text{(Sample Rate in Hz)} \times \text{(Segment Length in seconds)}\).
Given a source waveform \(x^{(i)}\), the objective is to generate a target waveform \(y^{(j)}\) where \(i \neq j \)).
We formulate this waveform conversion process as a mapping function \(f\), which generates the corresponding target waveform from the source waveform, denoted in \(f: \mathcal{X} \rightarrow \mathcal{Y}\).

\subsection{Approximation model}
Building on the previously defined process, we introduce our proposed framework: a two-stage model for multi-directional waveform conversion, detailed architecture is illustrated in Figure~\ref{fig:fig2}.
The initial component of the framework is the approximation model \(f_{\text{apx}}\), which generates a normalized waveform across modalities within a single model.
To enable effective training across multiple waveform types, we treat each type as a style domain and apply style-transfer techniques, specifically adaptive instance normalization (AdaIN).
Inspired by StarGANv2 style-transfer method~\cite{bib26}, the model learns to convert any input waveform to match the style of a target waveform domain.

\textbf{Model architecture and Style injection}

The approximation model consists of an encoder (\(E\)), a bottleneck (\(B\)), and a decoder (\(D\)).
The encoder employs two layers of 1-D convolution (CNN) followed by instance normalization to extract local waveform features, represented as \(h\).
Within the bottleneck, the feature vector \(h\) is processed by a Swin Transformer block, that captures global temporal dependencies and injects target waveform style via AdaIN, producing the style-conditioned feature vector \(\tilde{h}\).
Finally, the decoder generates the normalized target waveform \(\hat{y}_{\text{apx}}^{(j)}\) with AdaIN that adapts the style throughout the process.
The overall flow can be summarized as \( x^{(i)} \rightarrow h \rightarrow \tilde{h} \rightarrow \hat{y}_{\text{apx}}^{(j)} \), as described in Equation \ref{eq:model_flow}.

To inject target-specific style information into the bottleneck and decoder, the target type selector \(d^{(j)} \in \{0, 1\}^{|\mathcal{T}|}\)---one hot vector that indicates the type of target waveform \(j\)—  is transformed into a style-conditioned embedding vector \( s=\text{fc}(d^{(j)}) \in \mathbb{R}^{d_s} \) through a fully connected layer of dimensions  \(d_s\).
This embedding \(s\) is divided into two vectors \((\gamma, \beta)\) and each vector has \(s/2\) sizes of the feature dimensions. 
These two learnable vectors are applied in the AdaIN operation to adaptively shift and scale the characteristics $z$ -its mean $\mu(z)$ and standard deviation $\sigma(z)$- thus encoding the target waveform style as shown in Equation (\ref{eq:adain}).
AdaIN is applied throughout all layers of the SwinT\(_D\) block and the decoder \(D\), ensuring consistent style conditioning throughout the generation process.
In this way, the model does not require any specific modifications to each conversion task.

\textbf{Training with multiple waveform modalities} 

To enable simultaneous training across multiple waveform modalities, we pool ECG, PPG, and ABP waveforms into a same train set without explicit separation by modality for training the approximation model.
However, the differences in amplitude scales---such as the typically higher magnitudes of ABP compared to ECG or PPG---can cause the model to become biased toward these dominant modalities, leading to instability and degraded performance on others~\cite{bib62}.
To address this issue, we use waveforms with sample-wise normalization (local min-max normalization), guaranteeing that all samples have the same amplitude scale in the ranges 0-1.

During training, each mini-batch is constructed by sampling a source waveform segment of type \(i\), and then randomly selecting a different target waveform type \(j \ne i\), forming a source-to-target conversion pair \((i \rightarrow j)\) as defined in Equation~\ref{eq:trainset}.  
By including a diverse set of such conversion pairs in each mini-batch, the model learns to handle all possible translation directions across waveform types and to capture the underlying interrelationships among them.  
This training strategy allows the model to leverage all three waveform modalities during training while it generates any target modality from a single input waveform at inference time.
The model is trained to minimize the mean squared error (MSE) between the generated normalized waveform $\hat{y}_{\text{apx}}^{(j)}$ and the normalized ground truth waveform $y_{\text{apx}}^{(j)}$.
The overall model architecture and training objective are formalized as follows:

\begin{equation}
\begin{split}
E(x^{(i)}) &= h, \quad B(h, s) = \tilde{h}, \\
D(\tilde{h}, s) &= \hat{y}_{\text{apx}}^{(j)}, \quad 
s = [\gamma \ \| \ \beta], \\ 
z &= \mathrm{SwinT}_E(h)
\end{split}
\label{eq:model_flow}
\end{equation}

\begin{align}
\mathrm{AdaIN}(z, \gamma, \beta) &= (1 + \gamma) \cdot \frac{z - \mu(z)}{\sigma(z) + \varepsilon} + \beta \label{eq:adain} \\
\begin{split}
\mathcal{D}_{\text{train}} &= \big\{ (x^{(i)}, d^{(j)}, y^{(j)}) \mid\ \\ 
& \qquad i, j \in T, i \neq j \big\}
\end{split}
\label{eq:trainset} \\
\mathcal{L}_{\text{apx}} &= \frac{1}{L} \sum_{t=1}^{L} \left( \hat{y}_{\text{apx}}^{(j)} - y_{\text{apx}}^{(j)} \right)^2 \label{eq:loss}
\end{align}

\subsection{Refinement model}
While the approximation model generates the normalized waveform \(\hat{y}_{\text{apx},t}^{(j)}\), an additional refinement process is necessary to convert it into real-valued units necessary for downstream clinical decision-making. 
We introduce a refinement model \(f_{ref}\) that predicts amplitude information such as systolic blood pressure (SBP) and diastolic blood pressure (DBP), which correspond to the maximum and minimum values of the ABP waveform, and apply a linear transformation to scale the waveform in real-valued units (mmHg).
Note that since the dataset provides real-valued unit ground truths only for the ABP waveform, we apply the refinement procedure exclusively to ABP.

\begin{table*}[ht]
\renewcommand{\arraystretch}{1.25}
\centering
\resizebox{0.98\textwidth}{!}{
\begin{tabular}{cc|cc|ccc}
\hline
Dataset                  & Split                     & \multicolumn{2}{c|}{Train/Val} & \multicolumn{3}{c}{Test}    \\ \hline \hline
\multirow{3}{*}{PulseDB} & Number of Patient         & \multicolumn{2}{c|}{2,494}     & \multicolumn{3}{c}{279}     \\
                         & Samples for approximation & \multicolumn{2}{c|}{902,160}   & \multicolumn{3}{c}{111,535} \\
 &
  Samples for refinement &
  \begin{tabular}[c]{@{}c@{}}Pretrain- train\\ 721,728 (80\%)\end{tabular} &
  \begin{tabular}[c]{@{}c@{}}Pretrain-val\\ 180,432 (20\%)\end{tabular} &
  \begin{tabular}[c]{@{}c@{}}Finetune-train \\ 90,342 (81\%)\end{tabular} &
  \begin{tabular}[c]{@{}c@{}}Finetune-val\\ 10,038 (9\%)\end{tabular} &
  \begin{tabular}[c]{@{}c@{}}Finetune-test\\ 11,155 (10\%)\end{tabular} \\ \hline
\multirow{3}{*}{UCI}     & Number of Patient         & \multicolumn{2}{c|}{Unknown}       & \multicolumn{3}{c}{Unknown}     \\
                         & Samples for approximation  & \multicolumn{2}{c|}{140,552}   & \multicolumn{3}{c}{50,634}  \\
 &
  Samples for refinement &
  \begin{tabular}[c]{@{}c@{}}Pretrain- train\\ 112,441(80\%)\end{tabular} &
  \begin{tabular}[c]{@{}c@{}}Pretrain-val\\ 28,111(20\%)\end{tabular} &
  \begin{tabular}[c]{@{}c@{}}Finetune-train \\ 41,013 (81\%)\end{tabular} &
  \begin{tabular}[c]{@{}c@{}}Finetune-val\\ 4,557 (9\%)\end{tabular} &
  \begin{tabular}[c]{@{}c@{}}Finetune-test\\ 5,064 (10\%)\end{tabular} \\ \hline
\end{tabular}}

\caption{Summary of dataset splits for PulseDB and UCI.
The table reports the number of samples and patients arranged into training, validation, and testing sets.
The test set is calibration-free, meaning that test patients do not appear in the training set. 
To evaluate the benefit of pretraining, train/validation sets from the approximation model are used for refinement pretraining.
The original calibration-free test set is further split into training, validation, and testing subsets for finetuning.
This apply a calibration-based setting, where the same patient may appear in both the training and test sets.}
\label{tab:datasets}
\vskip -5pt
\end{table*}

\textbf{Multi-modal encoder}

First, the refinement model \( f_{\text{ref}} \) receives the source waveform \( x^{(i)} \) and the structured demographic information \( x_{\text{PI}} \) as inputs.
However, due to structural differences in demographic features across datasets (e.g. MIMIC-III waveform and VitalDB), schema-specific encoders must be manually designed, which is not scalable.
To address this, we convert the structured features into a linearized string format following text-based embedding studies~\cite{bib63, bib64}: “Age/Gender/Height/Weight/BMI”, using “/” as the delimiter.
This linearized text format enables the use of a tokenizer and text encoder as the patient information module \( E_{PI} \), where DistilBERT, a lightweight variant of BERT, encodes the tokenized subword units to produce the patient information embedding \( e_{PI} \).
This approach allows our framework to incorporate any structured patient information in a raw table format without requiring manual feature engineering or pre-processing.

Next, source waveform inputs \( x^{(i)} \) are passed through waveform-specific encoders \( E_W^{(i)} \), implemented using a shared PatchTSMixer backbone~\cite{bib34}.
The obtained embedddings \( e_W^{(i)} \) and \( e_{\text{PI}} \) are concatenated and passed to a regression module \( P^{(i)} \), which estimates the corresponding SBP \( \hat{y}_{\text{SBP}}\) and DBP \(\hat{y}_{\text{DBP}})\).
Note that this design allows the refinement model to flexibly operate with or without access to demographic information.

\textbf{Linear transformation}

To obtain an ABP waveform in real-valued units, we apply a linear transformation to the shape-only waveform generated by the approximation model, using the amplitude information predicted from patient-specific features.
Given the predicted \(\hat{y}_{\text{SBP}}\) and \(\hat{y}_{\text{DBP}}\) values from the multi-modal encoder, we perform a linear transformation to convert the normalized ABP waveform \(\hat{y}_{\text{apx}}^{ABP}\), generated by the approximation model, into a real-valued waveform in mmHg \(\hat{y}^{ABP}\).
This transformation maps the normalized waveform to the predicted range of DBP and SBP.
We follow the method proposed in NabNet~\cite{bib6}, and formulated as below:

\begin{align*}
e_W^{(i)} &= E_W^{(i)}(x^{(i)}), \quad e_{\text{PI}} = E_{\text{PI}}(x_{\text{PI}}) \\
\hat{y}_{\text{SBP}}, \hat{y}_{\text{DBP}} &= P^{(i)}([e_W^{(i)} ; e_{\text{PI}}]) \\
\hat{y}^{ABP} &= \hat{y}^{ABP}_{\text{apx}} \cdot (\hat{y}_{\text{SBP}} - \hat{y}_{\text{DBP}}) + \hat{y}_{\text{DBP}} \label{eq:nabnet}
\end{align*}

Note that, instead of directly predicting SBP and DBP in real-valued units (mmHg), we first estimate them in the globally min-max normalized ABP scale to ensure uniform target ranges and stabilize training.  
Using the predicted \(\hat{y}^{\text{norm}}_{\text{SBP}}\) and \(\hat{y}^{\text{norm}}_{\text{DBP}}\), we rescale the \textit{locally normalized waveform} to the \textit{globally normalized domain}, and then convert it to the final \textit{real-valued ABP waveform} \(\hat{y}^{\text{ABP}}\) using the global maximum and minimum values from the training set.

\textbf{Contrastive learning loss}

To accurately estimate the waveform in real-valued units, the refinement model is trained using multiple objectives that combine Mean Absolute Error(MAE) loss and weighted contrastive learning (WCL).
Contrastive learning enables models to learn an embedding space by minimizing the distance between positive pairs, samples that are semantically similar, while maximizing the distance between negative pairs. 
WCL extends this principle by introducing continuous similarity weights, derived from ground truth label similarities (e.g., SBP/DBP values, age, and gender), which are used to modulate the attraction between embeddings~\cite{bib31}.

In our refinement model, WCL guides the embedding spaces of waveform features and patient demographic information by aligning samples based on their clinical similarity.
For \( e_W^{(i)} \), a pair is considered similar if the MAE between their SBP and DBP values is less than or equal to 15\,mmHg. For \( e_{\text{PI}} \), a similar pair is defined as having an age difference less than 15 years and identical gender. 
Details of the similarity weights and their associated hyperparameters are described in Appendix~\ref{appendix:wcl}.

During training, we jointly optimize both ECG$\rightarrow$ABP and PPG$\rightarrow$ABP branches \emph{simultaneously}; hence the losses  are summed over $i\in\{\text{ECG},\text{PPG}\}$.
Although both ECG and PPG signals are jointly used during training, at inference time, the refinement model generates real-valued ABP waveforms from only a single input modality (either ECG or PPG).
Furthermore, contrastive learning loss can be used as self-supervised learning independently to pre-train the waveform and patient information encoders.
In this study, we apply pretraining with contrastive loss to initialize the encoders prior to downstream finetuning for both datasets.

\begin{align*}
\mathcal{L}_{\text{ref}} &= \mathcal{L}_{\text{MAE}} + \mathcal{L}_{\text{WCL}} \\
\begin{split}
\mathcal{L}_{\text{MAE}} &= \lambda\sum_{i \in \{\text{ECG}, \text{PPG}\}} \bigl(
|\hat{y}_{\text{SBP}\_(i)} - y_{\text{SBP}}|  \\
&\qquad \quad + \quad \qquad |\hat{y}_{\text{DBP}\_(i)} - y_{\text{DBP}}| \bigr)
\end{split} \\
\mathcal{L}_{\text{WCL}} &= \lambda_{1}\sum_{i \in \{\text{ECG}, \text{PPG}\}} \mathcal{L}_{e_W}^{(i)}
+ \lambda_{2}\, \mathcal{L}_{e_{\text{PI}}}
\end{align*}

\subsection{Implementation details}
\subsubsection{Data split}
We adopt a calibration-free data-splitting strategy for both the PulseDB and UCI datasets.
This approach splits the data at the patient level—ensuring that the train and test sets contain no overlapping patients—to assess whether the model generalizes to entirely unseen patients.
For each dataset, the training set is further divided into train and validation subsets in a 4:1 ratio.

Building on these splits, we further prepare a refinement model data split.
The original train and validation sets are used for self-supervised contrastive pretraining.
Separately, the original test set is further split into new train, validation, and test subsets to support finetuning under a calibration-based setting where the same patient may appear in both training and test sets.
This design allows us to evaluate whether inter-modality contrastive learning improves performance when used as a pretraining strategy.
The detailed split information is summarized in Table~\ref{tab:datasets}.

\subsubsection{Baselines}
Since there is no previous work that simultaneously handles multi-directional waveform conversion, we establish baselines using SOTA models from uni-directional waveform conversion tasks.
Each baseline, originally developed for a specific conversion direction, is adapted by training separate models one for each of the six waveform conversion pairs.
To fairly compare methods, we adjust the number of trainable parameters across baselines.
Detailed architectural descriptions and modifications of each baseline are provided below:

\begin{itemize}
    \item \textbf{NabNet}~\cite{bib6, bib23}: 
    A SOTA two-stage ABP conversion model with an approximation module (1-D CNN + attention-guided bi-CNN LSTM) and a refinement module (ShallowUnet + MLP).
    Approximation and refinement modules use 48 and 84 filter channels, respectively.
    
    \item \textbf{PPG2ABP}~\cite{bib4}: 
    A two-stage cascade 1-D U-Nets model for ABP conversion, where the approximation output from a deep-supervised U-Net feeds into a multi-resolution U-Net for refinement.
    Approximation and refinement modules use 96 filters channels and a scaling factor of 10, respectively.

    \item \textbf{PatchTST}~\cite{bib44}: 
     A two-stage transformer-based model using patch tokenization and temporal self-attention to jointly encode waveform and patient data. Previously adapted for ECG reconstruction~\cite{bib45} and BP prediction~\cite{bib31}.

    \item \textbf{P2E-WGAN}~\cite{bib19}: 
    A GAN for $\text{PPG}\rightarrow\text{ECG}$ synthesis, with CNN-based generator and discriminator.
    Approximation and refinement models (generator/discriminator) use 288/144 and 640/320 filter channels, respectively.
\end{itemize}

All methods, including the approximation model and the refinement model (both pretraining and finetuning), use identical hyperparameters: batch size of 2048,  learning rate of 1.00E-03, scheduler patience of 3, early stopping patience of 5, and a maximum of 30K training steps.
While all baselines adopt the same approximation approach as MD-ViSCo which generates a locally min-max normalized waveform, the refinement models are different for each.
NabNet and PatchTST follow the same methodology as MD-ViSCo, where the refinement model serves as a BP predictor and estimates SBP and DBP values. 
In contrast, PPG2ABP (two-stage) and P2E-WGAN (single-stage) models directly generate the ABP waveform.

\begin{figure*}[ht]
\centering
\centerline{\includegraphics[width=0.95\textwidth]{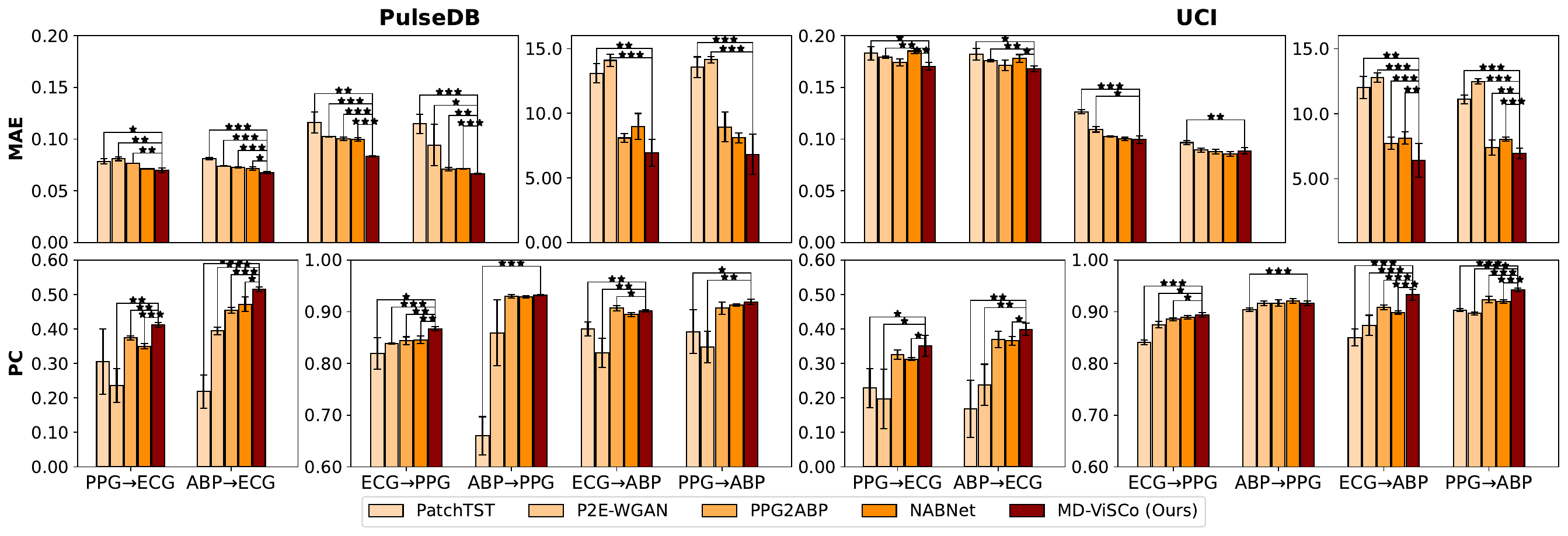}}
\caption{Comparison of MAE (↓) and PC (↑) from waveform conversion experiments on the PulseDB and UCI datasets across six waveform conversion directions (\textit{source waveform\,$\rightarrow$\, target waveform}).
Bar colors correspond to each method.
Error caps show $\pm$1 SD over five random seeds. 
MD-ViSCo (ours) method (dark red) consistently has shown better or comparable performance with baseline models (orange gradient).
Statistical significances between our method and others are assessed under an unpaired \emph{t}-test (*$p<0.05$, **$p<0.01$, ***$p<0.001$, n=5).}
\label{fig:fig3}
\vskip -5pt
\end{figure*}

\section{Experiment Results and Discussion}
We design four sets of experiments to evaluate MD-ViSCo against four baselines, assessing both waveform similarity and practical utility in health monitoring.
First, we conduct \emph{Multi-directional Waveform Conversion} experiment to evaluate our model’s capability to convert waveforms between different vital signs types (ECG, PPG, ABP).
Second, we perform a \emph{Physiological Feature Fidelity Evaluation} to assess the physiological relevance of converted waveforms by extracting morphology-based features and comparing them to those from the ground truth.
Third, we conduct an \emph{AAMI/BHS Standard Evaluation} to evaluate our ABP waveform conversions against established BP clinical standard~\cite{bib41, bib42}.
Lastly, we carry out ablation studies to validate the contributions of multi-directional training, WCL loss, and patient information to overall waveform generation performance.

\subsection{Multi-directional Waveform Conversion}
The goal of this experiment is to evaluate the performance of a single model across all conversion types within a multi-directional framework.
To ensure a fair comparison, each baseline is implemented as six separate uni-directional models, one per conversion direction.
As shown in Figure~\ref{fig:fig3}, we report the MAE and Pearson correlation (PC) between the converted and ground-truth waveforms across all six possible conversion directions, where MAE measures point-wise error and PC reflects how well the generated waveform preserves the overall shape and trend.
Lower MAE values indicate higher similarity, while more positive PC values reflect stronger linear relationship.
Note that ECG and PPG are evaluated in normalized units, whereas ABP is evaluated in real-valued units, resulting in task-specific Y-axis scales for MAE and PC.

Our MD-ViSCo consistently demonstrates the best performance in most conversion directions or shows comparable results to baseline models when the gap is not statistically significant.
Specifically, on the PulseDB dataset, it outperforms the average of all baseline models by 4.47 mmHg in MAE and 4\%P in PC for ABP waveform, by 0.007 MAE / 11.34\%P PC for ECG and 0.021 MAE / 5.88\%P PC for PPG generation.
In the UCI dataset, it showed improvements of 2.65 mmHg in MAE and 6\% in PC for ABP, 0.009 MAE / 9.93\%P PC for ECG, and 0.006 MAE / 1.16\%P PC PPG, again outperforming the baseline average.

PatchTST and P2E-WGAN demonstrate relatively lower performance: PatchTST is originally developed for general time series forecasting tasks, and P2E-WGAN suffers from unstable GAN training.
In contrast, PPG2ABP and NabNet show similar or lower performance compared to MD-ViSCo.
Unlike baseline methods that require six separate models, MD-ViSCo needs only a single model training.
This highlights that MD-ViSCo not only achieves strong performance but also reduces model maintenance and management costs, making it more efficient and practical for deployment in clinical settings.

\begin{figure*}[t]
\centering
\centerline{\includegraphics[width=0.98\textwidth]{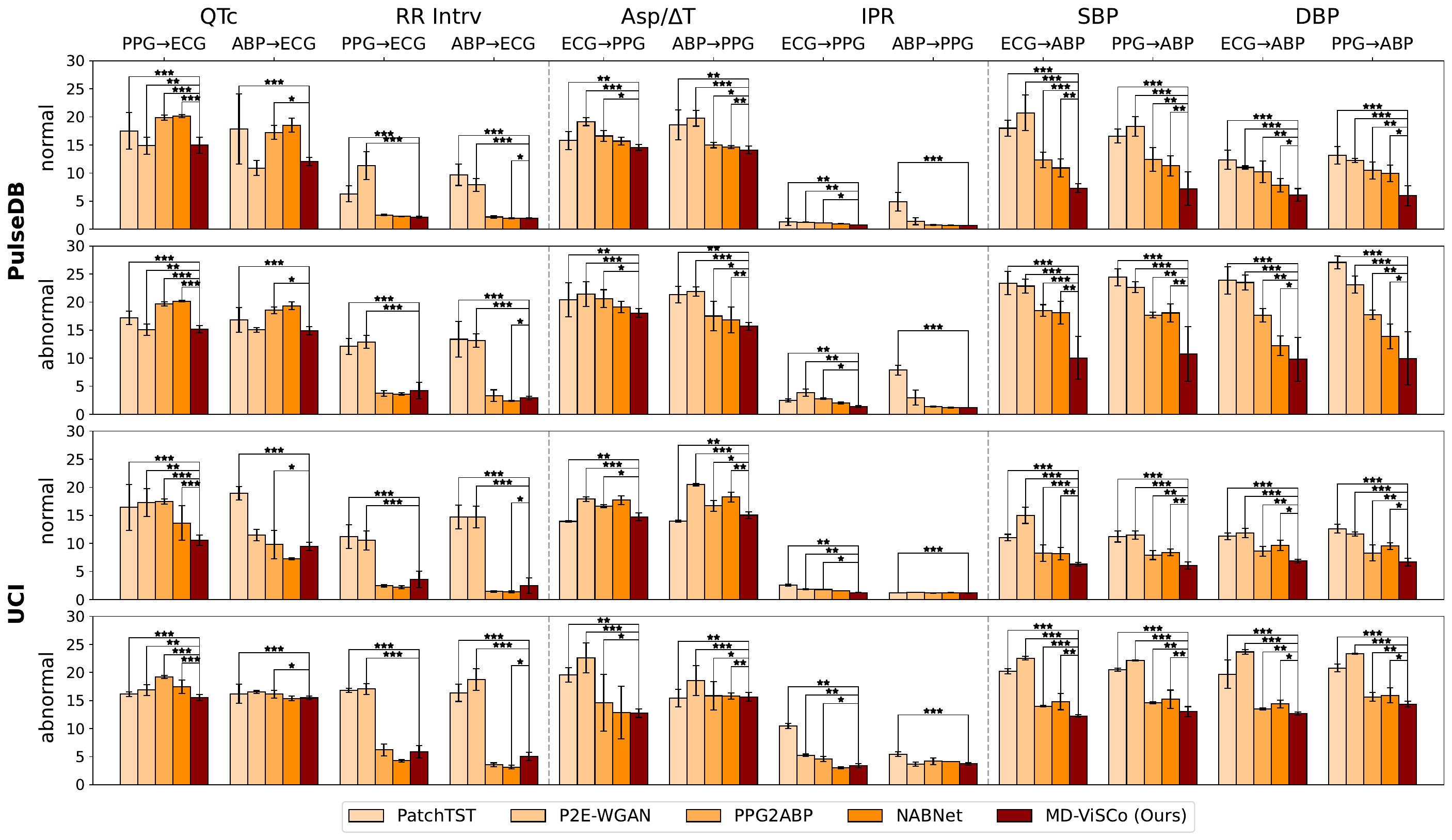}}
\caption{Comparison of relative mean error (\%) in physiological feature estimation across different waveform conversion directions and clinical subgroups.  
Each physiological feature is represented at the upper side, along with corresponding waveform conversion directions.
The y-axis label represents two datasets (PulseDB and UCI), and clinical subgroups(normal and abnormal).
Bar colors represent each method.
Error caps show $\pm$1 SD over five random seeds. 
Statistical significances between MD-ViSCo(ours) and others are assessed under an unpaired \emph{t}-test (*$p<0.05$, **$p<0.01$, ***$p<0.001$, n=5).}
\label{fig:feature}
\vskip -5pt
\end{figure*}

\subsection{Physiological Feature Fidelity Evaluation}
In addition to evaluating waveform similarity, we assess whether key morphological features, commonly used to derive physiological insight, are well preserved, assuming that features from well-generated waveforms should be similar to those from the ground truth.
Since clinical decision-making focuses on detecting abnormal cases, we group samples based on whether their ground-truth feature values lie within or outside the normal physiological range.
These normal ranges are defined using established clinical thresholds for each feature, as described below.
To account for features with varying value ranges, we measure relative error,  defined as the MAE divided by the absolute value of the feature, as the evaluation metric, shown in Figure~\ref{fig:feature}.

\subsubsection{ECG features}
We exclude P-peak-related features due to low sampling rates (125Hz) and instead focus on QRST-based features extracted using NeuroKit2 library~\cite{bib32}.

\textbf{The QTc interval [0.35-0.45s]} is the duration between the onset of the QRS complex and the end of the T-wave (QT interval), corrected for heart rate using Bazett's formula: $\text{QTc} = \frac{\text{QT}}{\sqrt{\text{RR}}}$.
Prolonged QTc ($>0.45$~s) may indicate congenital long QT syndrome or bradyarrhythmia; shortened QTc ($<0.35$~s) may suggest SQTS or hypercalcemia.
For QTc, the improvement is 3.9\%P on PulseDB and 4.8\%P on UCI over the baseline average in the normal range, and 2.9\%P on PulseDB and 0.4\%P on UCI in the abnormal range.

\textbf{RR interval [0.6-1.0s]} is a temporal feature that represents the time between two R peaks, divided by the sampling rate to compute beat-to-beat intervals in seconds (s).
Values $<0.6$~s may indicate sinus tachycardia, while values $>1.0$~s may reflect bradycardia or heart block.
For RR interval, which is the reciprocal of heart rate, it achieves a 3.4\%P improvement on PulseDB and 4.1\%P on UCI compared to the baseline average in the normal range, and a 4.3\%P improvement on PulseDB and 5.0\%P on UCI in the abnormal range.
\subsubsection{PPG features}
We evaluate two morphological features extracted using the pyPPG library~\cite{bib61}.
Due to the short duration of the waveform segments ($<10$ seconds), we exclude derivative-based features and focus on direct waveform features.

\textbf{Asp/$\Delta$T [2-3.5au/s]} is defined as the ratio of the systolic peak amplitude (Asp) to the time delay ($\Delta$T) between the waveform onset and the systolic peak, which is related to arterial stiffness.
Values below 2au/s may indicate hypotension or heart failure, while values above 3.5au/s may indicate hypertension or arteriosclerosis.
MD-ViSCo shows 2.3\%P improvement on PulseDB and 1.6\%P on UCI compared to the baseline average in the normal range, and 3.2\%P improvement on PulseDB and 3.0\%P on UCI in the abnormal range.

\textbf{Instantaneous Pulse Rate (IPR) [60-100BPM]} is reciprocal of the beat-to-beat interval.
Values below 60 BPM suggest bradycardia, while those above 100 BPM indicate tachycardia.
Our model achieves a 0.8\%P improvement on PulseDB and 0.3\%P on UCI over the baseline average in the normal range, and 1.7\%P improvement on PulseDB and 1.3\%P on UCI in the abnormal range. 

\begin{figure*}[] 
\centering
\centerline{\includegraphics[width=1\textwidth]{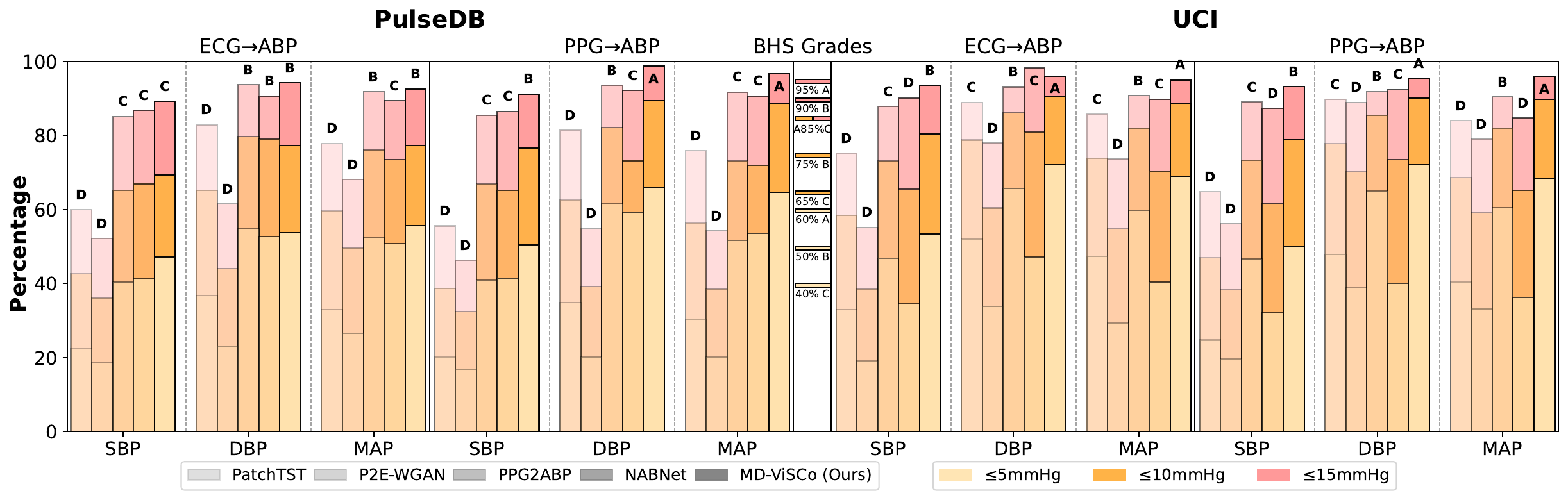}}
\caption{BHS-standard evaluation of generated ABP waveforms on the PulseDB and UCI test sets.  
Each bar shows the cumulative percentage of MAE values that fall within \(\le5\), \(\le10\), or \(\le15\)mmHg (peach, orange, red) for SBP, MAP, and DBP.
Bar opacity represents the methods.
The narrow center panel represents the official BHS grade cut-offs (A, B, C) for three thresholds.
For BHS grading, a method obtains a grade only if all three measures (SBP, DBP, MAP) meet the corresponding thresholds; otherwise, the overall grade is determined by the lowest among them(e.g., SBP C, MAP B, DBP A $\uparrow$ overall grade C).
Taller bars($\uparrow$) indicate better compliance.
}\label{fig:fig4}
\vskip -5pt
\end{figure*}

\begin{table*}[ht]
\setlength{\tabcolsep}{2.2pt}
\renewcommand{\arraystretch}{1.5}
\centering
\resizebox{1\textwidth}{!}{%
\begin{tabular}{c|c|cccccc|c|cccccc|c}
\hline
& & \multicolumn{6}{c|}{PPG$\rightarrow$ABP} & & \multicolumn{6}{c|}{ECG$\rightarrow$ABP} & \\ 
& & \multicolumn{2}{c}{SBP} & \multicolumn{2}{c}{MAP} & \multicolumn{2}{c|}{DBP} & AAMI & \multicolumn{2}{c}{SBP} & \multicolumn{2}{c}{MAP} & \multicolumn{2}{c|}{DBP} & AAMI \\ 
Dataset & Model & ME & Std & ME & Std & ME & Std &  & ME & Std & ME & Std & ME & Std & \\
\hline\hline
\multirow{5}{*}{PulseDB}
  & PatchTST & -0.60$_{\textit{\tiny1.03}}$ & 13.76$_{\textit{\tiny0.36}}$ & -0.16$_{\textit{\tiny1.72}}$ & 9.00$_{\textit{\tiny0.31}}$ & \textbf{0.06}$_{\textit{\tiny2.24}}$ & 7.92$_{\textit{\tiny0.32}}$ & Fail & -1.83$_{\textit{\tiny1.87}}$ & 13.03$_{\textit{\tiny0.68}}$ & -0.54$_{\textit{\tiny1.73}}$ & 8.73$_{\textit{\tiny0.43}}$ & \textbf{0.11}$_{\textit{\tiny1.77}}$ & 7.76$_{\textit{\tiny0.41}}$ & Fail \\
  & P2E-WGAN & -14.29$_{\textit{\tiny0.62}}$ & 14.25$_{\textit{\tiny0.21}}$ & -13.64$_{\textit{\tiny0.85}}$ & 10.34$_{\textit{\tiny0.20}}$ & -13.32$_{\textit{\tiny1.46}}$ & 11.01$_{\textit{\tiny0.49}}$ & Fail & 1.00$_{\textit{\tiny1.23}}$ & 14.29$_{\textit{\tiny0.39}}$ & -7.22$_{\textit{\tiny1.33}}$ & 9.88$_{\textit{\tiny0.19}}$ & -11.33$_{\textit{\tiny1.46}}$ & 9.62$_{\textit{\tiny0.31}}$ & Fail \\
  & PPG2ABP & 1.26$_{\textit{\tiny0.74}}$ & 8.28$_{\textit{\tiny1.32}}$ & -1.99$_{\textit{\tiny0.72}}$ & 6.39$_{\textit{\tiny0.91}}$ & -1.12$_{\textit{\tiny0.69}}$ & 6.17$_{\textit{\tiny0.87}}$ & Fail & 1.74$_{\textit{\tiny0.68}}$ & 9.18$_{\textit{\tiny0.40}}$ & -1.12$_{\textit{\tiny0.60}}$ & 6.12$_{\textit{\tiny0.44}}$ & -1.55$_{\textit{\tiny0.89}}$ & 6.34$_{\textit{\tiny0.95}}$ & Fail \\
  & NABNet & \textbf{0.60}$_{\textit{\tiny1.18}}$ & 10.27$_{\textit{\tiny0.97}}$ & \textbf{-0.05}$_{\textit{\tiny0.38}}$ & 7.39$_{\textit{\tiny0.36}}$ & -0.09$_{\textit{\tiny0.24}}$ & 6.43$_{\textit{\tiny0.29}}$ & Fail & -1.81$_{\textit{\tiny0.86}}$ & 10.81$_{\textit{\tiny0.52}}$ & -0.56$_{\textit{\tiny1.23}}$ & 7.93$_{\textit{\tiny0.72}}$ & -2.08$_{\textit{\tiny3.16}}$ & 6.79$_{\textit{\tiny0.77}}$ & Fail \\
  & \textbf{MD-ViSCo(ours)} & -0.89$_{\textit{\tiny0.43}}$ & \textbf{6.46}$_{\textit{\tiny2.46}}$ & -0.83$_{\textit{\tiny0.26}}$ & \textbf{4.03}$_{\textit{\tiny1.20}}$ & -1.04$_{\textit{\tiny0.63}}$ & \textbf{4.45}$_{\textit{\tiny1.29}}$ & \textbf{Pass} & \textbf{-0.09}$_{\textit{\tiny0.64}}$ & \textbf{7.09}$_{\textit{\tiny0.39}}$ & \textbf{-0.31}$_{\textit{\tiny0.69}}$ & \textbf{4.66}$_{\textit{\tiny0.52}}$ & -0.28$_{\textit{\tiny0.64}}$ & \textbf{4.38}$_{\textit{\tiny0.76}}$ & \textbf{Pass} \\
\hline
\multirow{5}{*}{UCI}
  & PatchTST & -1.19$_{\textit{\tiny1.11}}$ & 13.01$_{\textit{\tiny0.67}}$ & -0.94$_{\textit{\tiny0.59}}$ & 7.35$_{\textit{\tiny0.36}}$ & -0.82$_{\textit{\tiny0.76}}$ & 6.47$_{\textit{\tiny0.21}}$ & Fail & \textbf{-0.40}$_{\textit{\tiny1.16}}$ & 10.92$_{\textit{\tiny0.62}}$ & \textbf{0.30}$_{\textit{\tiny0.53}}$ & 6.46$_{\textit{\tiny0.31}}$ & 0.65$_{\textit{\tiny1.47}}$ & 5.89$_{\textit{\tiny0.23}}$ & Fail \\
  & P2E-WGAN & -3.43$_{\textit{\tiny2.17}}$ & 15.01$_{\textit{\tiny0.20}}$ & -1.41$_{\textit{\tiny1.10}}$ & 8.62$_{\textit{\tiny0.09}}$ & \textbf{-0.40}$_{\textit{\tiny1.19}}$ & 7.36$_{\textit{\tiny0.17}}$ & Fail & -2.82$_{\textit{\tiny2.73}}$ & 15.09$_{\textit{\tiny0.79}}$ & -5.46$_{\textit{\tiny4.72}}$ & 8.95$_{\textit{\tiny0.35}}$ & -4.77$_{\textit{\tiny1.09}}$ & 8.02$_{\textit{\tiny0.74}}$ & Fail \\
  & PPG2ABP & -1.49$_{\textit{\tiny0.81}}$ & 7.97$_{\textit{\tiny0.28}}$ & -0.84$_{\textit{\tiny0.72}}$ & 7.08$_{\textit{\tiny0.11}}$ & -0.51$_{\textit{\tiny1.86}}$ & 6.30$_{\textit{\tiny0.28}}$ & \textbf{Pass} & 1.78$_{\textit{\tiny0.68}}$ & 8.17$_{\textit{\tiny0.57}}$ & -1.29$_{\textit{\tiny1.54}}$ & 7.42$_{\textit{\tiny5.45}}$ & -1.82$_{\textit{\tiny0.68}}$ & 6.37$_{\textit{\tiny0.92}}$ & Fail \\
  & NABNet & -1.28$_{\textit{\tiny1.22}}$ & 9.87$_{\textit{\tiny1.02}}$ & -0.69$_{\textit{\tiny0.60}}$ & 5.49$_{\textit{\tiny0.56}}$ & -0.50$_{\textit{\tiny0.50}}$ & 5.15$_{\textit{\tiny0.56}}$ & Fail & -1.35$_{\textit{\tiny1.22}}$ & 8.61$_{\textit{\tiny0.98}}$ & -1.50$_{\textit{\tiny0.65}}$ & 7.27$_{\textit{\tiny0.46}}$ & -1.58$_{\textit{\tiny0.58}}$ & 6.95$_{\textit{\tiny0.25}}$ & Fail \\
  & \textbf{MD-ViSCo(ours)} & \textbf{1.15}$_{\textit{\tiny0.62}}$ & \textbf{7.86}$_{\textit{\tiny0.40}}$ & \textbf{-0.65}$_{\textit{\tiny0.61}}$ & \textbf{4.50}$_{\textit{\tiny0.17}}$ & -0.47$_{\textit{\tiny0.85}}$ & \textbf{4.77}$_{\textit{\tiny0.32}}$ & \textbf{Pass} & -1.47$_{\textit{\tiny0.83}}$ & \textbf{7.71}$_{\textit{\tiny0.77}}$ & -0.92$_{\textit{\tiny0.60}}$ & \textbf{5.60}$_{\textit{\tiny0.44}}$ & \textbf{-0.64}$_{\textit{\tiny0.51}}$ & \textbf{5.14}$_{\textit{\tiny0.36}}$ & \textbf{Pass} \\
\hline
\end{tabular}
}
\caption{AAMI compliance results. Mean Error (ME) and standard deviation (Std $\downarrow$) for SBP, MAP, and DBP estimation.
ME is computed by averaging errors across all detected SBP, MAP, or DBP points within a waveform; the corresponding standard deviation across those points is reported in Std. 
Each ME and Std value corresponds to the mean over five independent runs, with standard deviation across runs shown in italics. 
AAMI compliance results are shown on the right, and the best performance in each metric is highlighted in bold.}
\label{tab:aami_results}
\vskip -5pt
\end{table*}

\subsubsection{ABP features}:
\textbf{Systolic Blood Pressure (SBP) [90-130mmHg]} is ABP waveform maximum value (mmHg).
SBP below 90 mmHg may indicate hypotension, while values above 130 mmHg suggest hypertension.
MD-ViSCo reduces relative error by 4.5\%P on PulseDB and 31.9\%P on UCI in the normal range, and achieves larger improvements of 9.4\%P on PulseDB and 34.8\%P on UCI in the abnormal range.

\textbf{Diastolic Blood Pressure (DBP) [60-80mmHg]} is ABP waveform minimum value (mmHg). DBP below 60 mmHg may signal hypotension, while values above 80 mmHg can indicate hypertension.
MD-ViSCo shows a relative error improvement of 7.4\%P on PulseDB and 28.2\%P on UCI in the normal range, and further improvement of 9.6\%P on PulseDB and 31.1\%P on UCI in the abnormal range.

Across all features, the mean relative error tends to be slightly higher in the abnormal group, which reflects the inherent difficulty of modeling atypical patterns.
However, our model consistently maintains lower errors than baselines in both subgroups.
These results demonstrate that our model generates waveforms that preserve key diagnostic features compared to the baselines, even in abnormal cases.

\begin{table}[h]
\renewcommand{\arraystretch}{1.25}
\centering
\resizebox{\columnwidth}{!}{%
\begin{tabular}{|c|c|c|c|c|}
\hline
                               & ME(mmHg) & $\sigma$(mmHg) & Number of Subjects & AAMI Result \\ \hline
\multirow{2}{*}{AAMI Standard} & $\leq$5mmHg    & $\leq$8mmHg     & $\geq$85                 & Pass         \\ \cline{2-5} 
                               & $>$5mmHg    & $>$8mmHg     & $<$85                 & Fail         \\ \hline
\end{tabular}%
}
\caption{AAMI Standard requirements. AAMI requires ME, standard deviation, and the number of subject criteria to be fulfilled to pass the standard.}
\label{tab:tab6}
\vskip -5pt
\end{table}
% BHS Metric
\begin{table}[h]
\renewcommand{\arraystretch}{1.25}
\resizebox{\columnwidth}{!}{%
\centering
\begin{tabular}{|c|ccc|c|}
\hline
\multirow{2}{*}{}           & \multicolumn{3}{c|}{Cumulative MAE Percentage}       & \multirow{2}{*}{BHS Grade} \\ \cline{2-4}
 & \multicolumn{1}{c|}{$\leq$5mmHg} & \multicolumn{1}{c|}{$\leq$10mmHg} & $\leq$15mmHg &         \\ \hline
\multirow{3}{*}{BHS Metric} & \multicolumn{1}{c|}{60\%} & \multicolumn{1}{c|}{85\%} & 95\% & A                    \\ \cline{2-5} 
 & \multicolumn{1}{c|}{50\%}    & \multicolumn{1}{c|}{75\%}     & 90\%     & B \\ \cline{2-5} 
 & \multicolumn{1}{c|}{40\%}    & \multicolumn{1}{c|}{65\%}     & 85\%     & C \\ \cline{2-5} 
 & \multicolumn{1}{c|}{$\leq$40\%}    & \multicolumn{1}{c|}{$\leq$65\%}     & $\leq$85\%     & D \\ \hline
\end{tabular}%
}
\caption{BHS grade requirements. BHS grade is evaluated on the cumulative MAE percentage score.}
\label{tab:tab7}
\vskip -5pt
\end{table}

\subsection{AAMI/BHS Standard Evaluation}
To assess the clinical reliability of ABP waveforms generated by MD‑ViSCo, we evaluate whether they meet established clinical standards (AAMI/BHS) required for BP device approval.
Ensuring compliance with these standards demonstrates the potential for real-world application of our model, which are shown in Table~\ref{tab:aami_results} and Figure~\ref{fig:fig4} respectively.
Specifically, we compare SBP, DBP, and mean arterial pressure (MAP)\footnote{\(\text{MAP} = \frac{\text{SBP} + 2 \times \text{DBP}}{3}\)} from the generated ABP waveforms to ground-truth values.

\textbf{AAMI}: The AAMI standard which is a Pass/Fail assessment evaluate mean error (ME) and standard deviation(SD) on a dataset with more than 85 unique patients, as described in Table~\ref{tab:tab6}.
Our framework satisfies the AAMI criteria for both PPG$\rightarrow$ABP and ECG$\rightarrow$ABP conversions on both datasets with the lowest SD.
In contrast, except PPG2ABP which pass PPG$\rightarrow$ABP direction in UCI datasets, all baselines fail to meet the AAMI standard.
These results underscore the consistency and superiority of our framework, establishing it as the only model to satisfy the AAMI standard across all conversion tasks and datasets.

\textbf{BHS}: The BHS standard is detailed in Table~\ref{tab:tab7}, where specific MAE thresholds correspond to the grade.
The overall BHS grade is determined by the lowest grade among these nine individual evaluations. 
On the PulseDB dataset, our MD-ViSCo achieved grade B, outperforming all baseline methods:
PPG2ABP and NABNet attain grade C while PatchTST and P2E-WGAN get grade D in both directions.
On the UCI dataset, which contains a higher level of signal noise, our model is the only one that achieves grade B for both PPG$\rightarrow$ABP and ECG$\rightarrow$ABP directions, whereas PPG2ABP attains grade C, and all other baselines remain at grade D.

Taken together, these results demonstrate that MD-ViSCo consistently outperforms existing baselines in both AAMI and BHS compliances.
This confirms the strong potential of our model for real-world applications where waveform-based BP estimation is critical.

\begin{table}[t]
\setlength{\tabcolsep}{6pt}
\renewcommand{\arraystretch}{1.4}
\resizebox{\columnwidth}{!}{%
\centering
\begin{tabular}{l|cc|cc}
\hline
\multirow{2}{*}{Direction} & \multicolumn{2}{c|}{PulseDB} & \multicolumn{2}{c}{UCI} \\
                           & Uni & Multi & Uni & Multi \\ \hline
ECG$\rightarrow$PPG & \(0.093^{\ }_{\textit{\tiny0.001}}\) & \(0.083^{\ }_{\textit{\tiny0.001}}\) & \(0.102^{\ }_{\textit{\tiny0.001}}\) & \(0.098^{\ }_{\textit{\tiny0.003}}\) \\
ABP$\rightarrow$PPG & \(0.069^{\ }_{\textit{\tiny0.001}}\) & \(0.066^{\ }_{\textit{\tiny0.000}}\) & \(0.091^{\ }_{\textit{\tiny0.001}}\) & \(0.088^{\ }_{\textit{\tiny0.004}}\) \\
PPG$\rightarrow$ECG & \(0.075^{\ }_{\textit{\tiny0.001}}\) & \(0.071^{\ }_{\textit{\tiny0.001}}\) & \(0.172^{\ }_{\textit{\tiny0.001}}\) & \(0.171^{\ }_{\textit{\tiny0.004}}\) \\
ABP$\rightarrow$ECG & \(0.072^{\ }_{\textit{\tiny0.001}}\) & \(0.068^{\ }_{\textit{\tiny0.001}}\) & \(0.173^{\ }_{\textit{\tiny0.003}}\) & \(0.170^{\ }_{\textit{\tiny0.004}}\) \\
ECG$\rightarrow$ABP & \(7.23^{\ }_{\textit{\tiny1.00}}\)  & \(6.81^{\ }_{\textit{\tiny1.56}}\) & \(6.75^{\ }_{\textit{\tiny0.70}}\)  & \(6.41^{\ }_{\textit{\tiny0.54}}\) \\
PPG$\rightarrow$ABP & \(7.36^{\ }_{\textit{\tiny0.90}}\)  & \(6.95^{\ }_{\textit{\tiny1.04}}\) & \(7.14^{\ }_{\textit{\tiny0.50}}\)  & \(6.93^{\ }_{\textit{\tiny0.23}}\) \\ \hline
\end{tabular}%
}
\caption{Multi-directional versus uni-directional training comparison for six waveform conversion tasks. 
Each entry represents MAE($\downarrow$) averaged over five random-seed runs with standard deviation shown in italic subscript.}
\label{tab:waveform_mae_std}
\vskip -5pt
\end{table}
 
\begin{table}[h]
\setlength{\tabcolsep}{5pt}
\renewcommand{\arraystretch}{1.4}
\resizebox{\columnwidth}{!}{%
\centering
\begin{tabular}{cc|cc|cc}
\hline
\multicolumn{2}{c|}{Configure} & \multicolumn{2}{c|}{UCI} & \multicolumn{2}{c}{PulseDB} \\ \hline
WCL & PI & PPG$\rightarrow$ABP & ECG$\rightarrow$ABP & PPG$\rightarrow$ABP & ECG$\rightarrow$ABP \\
\hline\hline
N & N & 7.26$^{\ }_{\textit{\tiny0.36}}$ & 7.01$^{\ }_{\textit{\tiny0.39}}$ & 8.50$^{\ }_{\textit{\tiny1.30}}$ & 8.03$^{\ }_{\textit{\tiny1.16}}$ \\
Y & N & 6.93$^{\ }_{\textit{\tiny0.23}}$ & 6.41$^{\ }_{\textit{\tiny0.54}}$ & 7.83$^{\ }_{\textit{\tiny0.62}}$ & 7.53$^{\ }_{\textit{\tiny1.32}}$ \\
N & Y & -- & -- & 7.60$^{\ }_{\textit{\tiny1.21}}$ & 7.26$^{\ }_{\textit{\tiny0.32}}$ \\
Y & Y & -- & -- & 6.94$^{\ }_{\textit{\tiny1.41}}$ & 6.81$^{\ }_{\textit{\tiny1.04}}$ \\
\hline
\end{tabular}%
}
\caption{Ablation study of contrastive learning and patient information.
Each entry represents MAE($\downarrow$) averaged over five seeds in mmHg, with the tiny italic subscript indicating standard deviation. WCL indicates whether \(\mathcal{L}_\text{WCL}\) is added to the training objective. 
PI specifies whether \(x_{\text{PI}}\) is utilized by the refinement model. 
As the UCI dataset does not provide patient demographics, PI variants are not applicable.}
\label{tab:wcl_ablation}
\vskip -5pt
\end{table}

\subsection{Ablation Study}
\textbf{Uni-directional Waveform Conversion:}

We conduct an ablation study comparing multi-directional and uni-directional training, aiming to demonstrate that a single model trained across multiple waveform directions outperforms separately trained models for individual conversion tasks.
Summarized in Table~\ref{tab:waveform_mae_std}, we report the MAE and SD ($\sigma$) across the six waveform conversion directions.
Multi-directional training consistently outperforms uni-directional models across all waveform conversion tasks.
We observe average MAE reductions of 5.4\%P for ECG generation, 3.1\%P for PPG generation, and 4.0\%P for ABP generation, demonstrating consistent gains across waveform types without requiring task-specific training.
These results exhibit that multi-directional joint training improves overall performance and generalization by learning shared representations across waveform conversion tasks.

\textbf{WCL loss and Patient Information:}

To assess the individual contribution of WCL loss and patient information (PI) in the refinement model, we conduct an ablation study by selectively removing each component.  
Specifically, we compare the full model against variants where WCL loss, PI, or both are removed, allowing us to isolate the effect of each component.
Since PI is not available in UCI, configurations involving the PI component are used only with PulseDB.

As shown in Table~\ref{tab:wcl_ablation}, WCL consistently improves performance on both datasets, while PI provides additional gains on PulseDB.
Specifically, adding WCL reduced MAE by 0.33mmHg on UCI and 0.67mmHg on PulseDB in the PPG$\rightarrow$ABP task.
PI alone contributes an additional 0.90mmHg reduction, and combining both WCL and PI yield a total improvement of 1.56mmHg over the baseline.
These results indicate that WCL and PI offer complementary benefits for ABP waveform generation by aligning embeddings and enabling patient-specific calibration.
We adopt this configuration as the default for our refinement model.

\section{Conclusion}
\label{sec:conclusion}
In this study, we propose MD-ViSCo that enables efficient multi-directional conversion among ECG, PPG, and ABP waveforms.
Our model not only demonstrates superior or comparable performance to previous uni-directional baselines, but also validates health monitoring applicability through physiological feature fidelity and clinical compliance with BP medical device standards.
We believe this framework will pave the way for scalable waveform conversion solutions in the healthcare field.

Although our framework is promising, several limitations still exist.
It requires an additional refinement step to recover real-valued waveforms, which may involve task- or dataset-specific finetuning.
Also, the current design excludes other valuable modalities such as clinical notes or time-series electronic health records (EHR).
To address these limitations, future research will explore a language-model-based architecture that learns to generate real-valued waveforms directly from textual representations of vital sign waveforms.

\textbf{Acknowledgment}
This work was supported by the Institute of Information \& Communications Technology Planning \& Evaluation (IITP) grant (No.~RS-2019-II190075) funded by the Korea government (MSIT).

\section{Appendix}
\label{sec:appendix}

\subsection{Data Pre-processing} \label{appendix:dataset}
\subsubsection{Waveforms pre-processing}

\textbf{Local min-max normalization} is applied sample-wise for the approximation model, since ABP is provided in its original physical units (mmHg).
\textbf{Global Min-Max normalization} ensures that the model’s output can be linearly transformed back into real-valued unit (mmHg) using the minimum DBP and maximum SBP from the training set as fixed global bounds, making the output directly comparable to true the ABP.
This global normalization is applied to ABP waveforms, SBP, DBP, and MAP values in both the UCI and PulseDB datasets, using fixed bounds of 50–189.98 mmHg for UCI and 2.34–286.58 mmHg for PulseDB.
\textbf{Zero-centering} is applied only to the PulseDB dataset after local min-max normalization and is not used for globally min-max normalized ABP waveforms, in order to preserve their real-valued amplitude in mmHg.
This step mostly corrects ECG baseline offset, often caused by device-specific differences, which is more noticeable in PulseDB due to its combination of data from MIMIC-III and VitalDB with varying acquisition settings.
\textbf{Zero-padding} ensures waveform length multiples of 2. For PulseDB waveforms originally consisting of 1250 samples (10 seconds at 125 Hz), we apply 15-sample symmetric zero-padding on both ends, resulting in a final length of 1280 samples.

\subsubsection{Patient information pre-processing}
Patient information includes age (years), gender (Male/Female), height (cm), weight (kg), and body mass index (BMI, kg/m²); missing values for BMI, height, and weight in the MIMIC-III patient table are left as null.

\subsection{Model Architecture Detail}
All other details not mentioned in this section follow the default settings of the original implementations~\cite{bib25,bib26,bib30,bib34,bib60}
\subsubsection{Approximation model architecture}
The U-Net encoder starts with a 1-D CNN (kernel size 3, stride 1, filter channels 64).
The encoder downsamples using a residual block (instance normalization, LeakyReLU), followed by parallel max-pooling and strided CNN (kernel size 2, stride 2) paths.
Their outputs are concatenated and passed through another residual block. 
SwinT encoder includes patch embedding (output channels 256) and Swin Transformer blocks (window size 4, 32 attention heads per stage), both modified for taking 1-D waveforms. 
SwinT decoder consists of Swin Transformer blocks with AdaIN (style vector dimension 64) and patch-expanding operations.
The U-Net decoder is composed of two AdaIN-residual blocks and upsample the output using interpolation (scale factor 2) and transposed CNN.
The 1x1 CNN in the last layer reduces the 64-channel feature map to a single-channel waveform.

\subsubsection{Refinement model architecture}
Waveforms are encoded using \textit{PatchTSMixer} (hidden size 64, 15 layers, expansion factor 5, patch length 4), while patient demographic information is processed using the tokenizer and encoder of \textit{DistilBERT} \texttt{base-uncased} (6 layers, 12 heads, hidden size 768). 
Resulting embedding feature is projected to a 512‑dimensional space through a two‑layer MLP (GeLU, dropout 0.1), yielding a $512$‑dimensional representation.
The BP regression module takes the combined 512-dimensional waveform and patient embeddings as input and uses a PatchTSMixer followed by two-layer MLP heads.

\subsection{Detailed information for WCL}\label{appendix:wcl}
\textbf{BP-based similarity weights:}

{\small
\begin{equation}
\begin{split}
S^*_{ij} = \frac{1}{2} \bigg( 
&\exp\left(-\frac{|SBP_i - SBP_j|}{\tau_s}\right) \\
+ \quad &\exp\left(-\frac{|DBP_i - DBP_j|}{\tau_s}\right) 
\bigg)
\end{split}
\end{equation}
}

\textbf{Age-based similarity weights:}

{\small
\begin{equation}
\begin{split}
S^{*\,\text{age}}_{ij} &= \exp\left(-\frac{|age_i - age_j|}{\tau_s}\right), \\
\tau_s &= 4,\quad T_s = 0.0235
\end{split}
\end{equation}
}

\textbf{Gender-based similarity weights:}

{\small
\begin{equation}
\begin{split}
S^{*\,\text{gender}}_{ij} &=
\begin{cases} 
1 & \text{if } \text{gender}_i = \text{gender}_j \\
0 & \text{otherwise} 
\end{cases}, \\
\tau_s &= 1,\quad T_s = 1
\end{split}
\end{equation}
}

\begin{table}[H]
\resizebox{\columnwidth}{!}{%
\centering
\begin{tabular}{lll}
\textbf{Parameter} & \textbf{Description} & \textbf{Value} \\
$\lambda_{\text{MAE}}$ & MAE weight for BP regression loss & 0.001 \\
$\lambda_1$ & Weight for waveform embedding WCL & 0.01 \\
$\lambda$ & Weight for patient embedding WCL components & 0.01 \\
$\tau_s$ (waveform) & Similarity decay for SBP/DBP differences & 4 \\
$T_s$ (waveform) & Threshold to retain waveform similarity & 0.0235 \\
$\tau_s$ (age) & Similarity decay for age differences & 4 \\
$T_s$ (age) & Threshold to retain age similarity & 0.0235 \\
$\tau_s$ (gender) & Decay for gender similarity (binary) & 1 \\
$T_s$ (gender) & Threshold for gender similarity (binary) & 1 \\
$\tau_w$ & Temperature for softmax normalization & 4 \\
\bottomrule
\end{tabular}%
}
\caption{WCL hyperparameters used in training $\mathbf{M_{ref}}$.}
\label{tab:wcl_hyperparams}
\end{table}

\begin{thebibliography}{00}
\bibitem{bib46}\label{bib46} A. J. E. Seely and K. Newman and R. Ramchandani and C. Herry and N. Scales and N. Hudek and J. Brehaut and D. Jones and T. Ramsay and D. Barnaby and S. Fernando and J. Perry and S. Dhanani and K. E. A. Burns, ``Roadmap for the evolution of monitoring: developing and evaluating waveform-based variability-derived artificial intelligence-powered predictive clinical decision support software tools,'' {\it Critical Care}, vol. 28, no. 1, pp. 404, 2024, {{\it Springer Nature}}, doi: {10.1186/s13054-024-05140-6}.

\bibitem{bib48}\label{bib48} M. Mollura and L.-W. H. Lehman and R. G. Mark and R. Barbieri, ``A novel artificial intelligence based intensive care unit monitoring system: using physiological waveforms to identify sepsis,'' {\it Philosophical Transactions of the Royal Society A: Mathematical, Physical and Engineering Sciences}, vol. 379, no. 2212, pp. 20200252, 2021, doi: {10.1098/rsta.2020.0252}.

\bibitem{bib50}\label{bib50} B. Rowland, A. Saha, V. Motamedi, R. Bundy, S. Winsor, D. McNavish, W. Lippert, and A. K. Khanna, 
``Impact on Patient Outcomes of Continuous Vital Sign Monitoring on Medical Wards: Propensity-Matched Analysis,'' 
\emph{J. Med. Internet Res.}, vol. 27, p. e66347, Mar. 11, 2025, 
doi: 10.2196/66347.

\bibitem{bib21}\label{bib21} X. Tian and Q. Zhu and Y. Li and M. Wu, ``Cross-Domain Joint Dictionary Learning for ECG Inference From PPG,'' {\it IEEE Internet of Things Journal}, vol. 10, no. 9, pp. 8140--8154, 2023, doi: 10.1109/JIOT.2022.3231862.

\bibitem{bib54}\label{bib54} A. Dash and N. Ghosh and A. Patra and A. D. Choudhury, ``Estimation of Arterial Blood Pressure Waveform from Photoplethysmogram Signal using Linear Transfer Function Approach,'' in {\it Proceedings of the 2020 42nd Annual International Conference of the IEEE Engineering in Medicine \& Biology Society (EMBC)}, Montreal, QC, Canada, 2020, pp. 2691--2694, doi: 10.1109/EMBC44109.2020.9175696.

\bibitem{bib13}\label{bib13} T. Athaya and S. Choi, ``A review of noninvasive methodologies to estimate the blood pressure waveform,'' {\it Sensors}, vol. 22, no. 10, pp. 3953, 2022, {{\it MDPI}} doi: 10.3390/s22103953.

\bibitem{bib56}\label{bib56} B. V. Scheer and A. Perel and U. J. Pfeiffer, ``Clinical review: Complications and risk factors of peripheral arterial catheters used for haemodynamic monitoring in anaesthesia and intensive care medicine,'' {\it Critical Care}, vol. 6, no. 3, pp. 199, 2002, {{\it Springer Nature}} doi: 10.1186/cc1489.

\bibitem{bib2}\label{bib2} T. Athaya and S. Choi, ``An estimation method of continuous non-invasive arterial blood pressure waveform Using photoplethysmography: a U-Net architecture-based approach,'' {\it Sensors}, vol. 21, no. 5, pp. 1867, 2021, {{\it MDPI}} doi: 10.3390/s21051867.

\bibitem{bib37}\label{bib37} G. Nie, J. Zhu, G. Tang, D. Zhang, S. Geng, Q. Zhao, \emph{et al.}, 
``A review of deep learning methods for photoplethysmography data,'' 
\emph{arXiv preprint arXiv:2401.12783}, 2024. [Online]. Available: 10.48550/arXiv.2401.12783

\bibitem{bib53}\label{bib53} A. Sullivan and H. Xia and J. McBride and X. Zhao, ``Reconstruction of missing physiological signals using artificial neural networks,'' in {\it Proceedings of the 2010 Computing in Cardiology}, Belfast, UK, 2010, pp. 317--320.

\bibitem{bib52}\label{bib52} X. Hu, ``Foundation models for physiological data, how to develop them, and what to expect of them,'' {\it Physiological Measurement}, vol. 45, no. 2, pp. 020301, 2024, {{\it IOP Publishing}} doi: 10.1088/1361-6579/ad252f.

\bibitem{bib7}\label{bib7}  G. Zhang and D. Choi and D. E. Kim and J. Jung, Reconstruction of ABP waveform from the ECG or PPG signals using enhanced 1-D UNetwork. Presented at 2023 IEEE/ACIS 23rd International Conference on Computer and Information Science (ICIS). [Online]. Available: {10.1109/ICIS57766.2023.10210223}

\bibitem{bib36}\label{bib36} J. Pan and L. Liang and Y. Liang and Q. Tang and Z. Chen and J. Zhu, ``Robust modelling of arterial blood pressure reconstruction from photoplethysmography,'' {\it Scientific Reports}, vol. 14, no. 1, pp. 30333, 2024, {{\it MDPI}} doi: 10.1038/s41598-024-82026-1.

\bibitem{bib35}\label{bib35} M. Kachuee, M. M. Kiani, H. Mohammadzade, and M. Shabany, 
``Cuff-less high-accuracy calibration-free blood pressure estimation using pulse transit time,'' 
in \emph{Proc. 2015 IEEE Int. Symp. Circuits Syst. (ISCAS)}, Long Beach, CA, USA, 2015, pp. 1006--1009, doi: 10.1109/ISCAS.2015.7168806.

\bibitem{bib24}\label{bib24} W. Wang and P. Mohseni and K. L. Kilgore and L. Najafizadeh, ``PulseDB: A large, cleaned dataset based on MIMIC-III and VitalDB for benchmarking cuff-less blood pressure estimation methods,'' {\it Frontiers in Digital Health}, vol. 4, 2023, doi: 10.3389/fdgth.2022.1090854.

\bibitem{bib4}\label{bib4} N. Ibtehaz and S. Mahmud and M. E. H. Chowdhury and A. Khandakar and M. Salman Khan and M. A. Ayari and A. M. Tahir and M. S. Rahman, ``PPG2ABP: translating photoplethysmogram (PPG) signals to arterial blood pressure (ABP) waveforms,'' {\it Bioengineering}, vol. 11, no. 9, pp. 692, 2022, {{\it MDPI}} doi: 10.3390/bioengineering9110692.

\bibitem{bib6}\label{bib6} S. Mahmud, N. Ibtehaz, A. Khandakar, M. S. Rahman, A. J. R. Gonzales, T. Rahman, \emph{et al.}, 
``NABNet: A nested attention-guided BiConvLSTM network for a robust prediction of blood pressure components from reconstructed arterial blood pressure waveforms using PPG and ECG signals,'' 
\emph{Biomed. Signal Process. Control}, vol. 79, p. 104247, 2023, doi: 10.1016/j.bspc.2022.104247.

\bibitem{bib18}\label{bib18} Q. Zhu and X. Tian and C.-W. Wong and M. Wu, ``ECG Reconstruction via PPG: A Pilot Study,'' in {\it 2019 IEEE EMBS International Conference on Biomedical \& Health Informatics (BHI),}Chicago, IL, USA,  2019, pp. 1--4, doi: 10.1109/BHI.2019.8834612.

\bibitem{bib22}\label{bib22} Q. Zhu, X. Tian, C.-W. Wong, and M. Wu, 
``Learning your heart actions from pulse: ECG waveform reconstruction from PPG,'' 
\emph{IEEE Internet of Things Journal}, vol. 8, no. 23, pp. 16734--16748, Dec. 2021, doi: 10.1109/JIOT.2021.3097946.

\bibitem{bib9}\label{bib9}  K. Rishi Vardhan and S. Vedanth and G. Poojah and K. Abhishek and M. Nitish Kumar and V. Vijayaraghavan, BP-Net: efficient deep Learning for continuous arterial blood pressure estimation using photoplethysmogram. Presented at 2021 20th IEEE International Conference on Machine Learning and Applications (ICMLA). [Online]. Available: {10.1109/ICMLA52953.2021.00241}

\bibitem{bib12}\label{bib12} Z. Li and W. He, ``A continuous blood pressure estimation method using photoplethysmography by GRNN-based model,'' {\it Sensors}, vol. 21, no. 21, pp. 7207, 2021, {{\it MDPI}} doi: 10.3390/s21217207.

\bibitem{bib15}\label{bib15} S. Dasgupta and S. Das and U. Bhattacharya, ``CardioGAN: An Attention-based Generative Adversarial Network for Generation of Electrocardiograms,'' in {\it 2020 25th International Conference on Pattern Recognition (ICPR),}Milan, Italy,  2021,
pp. 3193--3200, doi: 10.1109/ICPR48806.2021.9412905.

\bibitem{bib19}\label{bib19} K. Vo, E. K. Naeini, A. Naderi, D. Jilani, A. M. Rahmani, N. Dutt, \emph{et al.}, 
``P2E-WGAN: ECG waveform synthesis from PPG with conditional Wasserstein generative adversarial networks,'' 
in \emph{Proc. 36th Annu. ACM Symp. Appl. Comput.}, Virtual Event, Republic of Korea, 2021, pp. 1030--1036, doi: 10.1145/3412841.3441979.

\bibitem{bib16}\label{bib16} Q. Tang and Z. Chen and R. Ward and C. Menon and M. Elgendi, ``PPG2ECGps: An End-to-End Subject-Specific Deep Neural Network Model for Electrocardiogram Reconstruction from Photoplethysmography Signals without Pulse Arrival Time Adjustments,'' {\it Bioengineering}., vol. 10, no. 6, pp. 630, 2023, doi: 10.3390/bioengineering10060630.

\bibitem{bib20}\label{bib20} Q. Tang and Z. Chen and Y. Guo and Y. Liang and R. Ward and C. Menon and M. Elgendi, ``Robust Reconstruction of Electrocardiogram Using Photoplethysmography: A Subject-Based Model,'' {\it Frontiers in Physiology}, vol. 13, 2022, doi: 10.3389/fphys.2022.859763.

\bibitem{bib14}\label{bib14} N. Aguirre and E. Grall-Maës and L. J. Cymberknop and R. L. Armentano, ``Blood pressure morphology assessment from photoplethysmogram and demographic information using deep learning with attention mechanism,'' {\it Sensors}, vol. 21, no. 6, pp. 2167, 2021, {{\it MDPI}} doi: 10.3390/s21062167.

\bibitem{bib8}\label{bib8} B. L. Hill and N. Rakocz and {\'A}. Rudas and J. N. Chiang and S. Wang and I. Hofer and M. Cannesson and E. Halperin, ``Imputation of the continuous arterial line blood pressure waveform from non-invasive measurements using deep learning,'' {\it Scientific Reports}, vol. 11, no. 1, pp. 15755, 2021, {{\it MDPI}} doi: 10.1038/s41598-021-94913-y.

\bibitem{bib38}\label{bib38} H.-C. Lee and Y. Park and S. B. Yoon and S. M. Yang and D. Park and C.-W. Jung, ``VitalDB, a high-fidelity multi-parameter vital signs database in surgical patients,'' {\it Scientific Data}, vol. 9, no. 1, pp. 279, 2022, {{\it Nature}} doi: 10.1038/s41597-022-01411-5.

\bibitem{bib39}\label{bib39} B. Moody and G. Moody and M. Villarroel and G. D. Clifford and I. Silva, ``MIMIC-III Waveform Database Matched Subset (version 1.0),'' {\it PhysioNet}, 2020. [Online]. Available: https://doi.org/10.13026/c2294b

\bibitem{bib26}\label{bib26} Y. Choi and Y. Uh and J. Yoo and J. Ha, ``StarGAN v2: Diverse Image Synthesis for Multiple Domains,'' in {\it 2020 IEEE/CVF Conference on Computer Vision and Pattern Recognition (CVPR),}Seattle, WA, USA,  2020, pp. 8185--8194, doi: 10.1109/CVPR42600.2020.00821.

\bibitem{bib62}\label{bib62} E. S. Lubana, R. P. Dick, and H. Tanaka, 
``Beyond BatchNorm: Towards a Unified Understanding of Normalization in Deep Learning,'' 
in \emph{Proc. 35th Int. Conf. Neural Inf. Process. Syst. (NeurIPS)}, Red Hook, NY, USA: Curran Associates Inc., 2021, Art. no. 365, pp. 1--14. [Online]. Available: \url{https://papers.nips.cc/paper_files/paper/2021/file/e317d34eb9d99eecfa2e41ce43f1412e-Paper.pdf}

\bibitem{bib63}\label{bib63} K. Hur, J. Lee, J. Oh, W. Price, Y. Kim, and E. Choi, 
``Unifying heterogeneous electronic health records systems via text-based code embedding,'' 
in \emph{Proc. Conf. Health, Inference, and Learn. (CHIL)}, vol. 174, PMLR, pp.~183--203, Apr. 2022. [Online]. Available: https://proceedings.mlr.press/v174/hur22a.html

\bibitem{bib64}\label{bib64} K. Hur, J. Oh, J. Kim, J. Kim, M. J. Lee, E. Cho, \emph{et al.}, 
``GenHPF: General healthcare predictive framework for multi-task multi-source learning,'' 
\emph{IEEE J. Biomed. Health Inform.}, vol. 28, no. 1, pp.~502--513, Jan. 2024, doi: 10.1109/JBHI.2023.3327951.

\bibitem{bib31}\label{bib31} S. B. N., R. S. Srinivasa, Y. M. Saidutta, J. Cho, C.-H. Lee, C. Yang, \emph{et al.}, 
``End-to-end personalized cuff-less blood pressure monitoring using ECG and PPG signals,'' 
in \emph{Proc. IEEE Int. Conf. Acoust., Speech Signal Process. (ICASSP)}, Seoul, South Korea, 2024, pp.~2101--2105, doi: 10.1109/ICASSP48485.2024.10445970.

\bibitem{bib34}\label{bib34} V. Ekambaram, A. Jati, N. Nguyen, P. Sinthong, and J. Kalagnanam, ``TSMixer: Lightweight MLP-Mixer Model for Multivariate Time Series Forecasting,'' in {\it Proceedings of the 29th ACM SIGKDD Conference on Knowledge Discovery and Data Mining,}Long Beach, CA, USA,  2023, pp. 459-–469, doi: 10.1145/3580305.3599533.

\bibitem{bib23}\label{bib23} S. Mahmud and N. Ibtehaz and A. Khandakar and A. M. Tahir and T. Rahman and K. R. Islam and M. S. Hossain and M. S. Rahman and F. Musharavati and M. A. Ayari and M. T. Islam and M. E. H. Chowdhury, ``A Shallow U-Net Architecture for Reliably Predicting Blood Pressure (BP) from Photoplethysmogram (PPG) and Electrocardiogram (ECG) Signals,'' {\it Sensors}, vol. 22, no. 3, pp. 919, 2022, doi: 10.3390/s22030919.

\bibitem{bib44}\label{bib44} Y. Nie and N. H. Nguyen and P. Sinthong and J. Kalagnanam, ``A Time Series is Worth 64 Words: Long-term Forecasting with Transformers,'' in {\it Proceedings of the Eleventh International Conference on Learning Representations (ICLR)}, 2023. [Online]. Available: \url{https://openreview.net/forum?id=Jbdc0vTOcol}

\bibitem{bib45}\label{bib45} Y. Li, M. Wang, M. Guan, C. Lu, Z. Li, and T. Chen, 
``SAPTSTA-AnoECG: A PatchTST-based ECG anomaly detection method with subtractive attention and data augmentation,'' 
\emph{Appl. Intell.}, vol. 55, no. 3, Art. no. 184, 2025, doi: 10.1007/s10489-024-05881-5.

\bibitem{bib41}\label{bib41} E. O'Brien, J. Petrie, W. Littler, M. de Swiet, P. L. Padfield, K. O'Malley, M. Jamieson, D. Altman, M. Bland, and N. Atkins, 
``The British Hypertension Society protocol for the evaluation of automated and semi-automated blood pressure measuring devices with special reference to ambulatory systems,'' 
\emph{J. Hypertens.}, vol. 8, no. 7, pp.~607--619, 1990, doi: 10.1097/00004872-199007000-00004.

\bibitem{bib42}\label{bib42} W. B. White and A. S. Berson and C. Robbins and M. J. Jamieson and L. M. Prisant and E. Roccella and S. G. Sheps, ``National standard for measurement of resting and ambulatory blood pressures with automated sphygmomanometers,'' {\it Hypertension}, vol. 21, no. 4, pp. 504--509, 1993, doi: 10.1161/01.HYP.21.4.504.

\bibitem{bib32}\label{bib32} D. Makowski, T. Pham, Z. J. Lau, J. C. Brammer, F. Lespinasse, H. Pham, C. Schölzel, and S. A. Chen, 
``NeuroKit2: A Python toolbox for neurophysiological signal processing,'' 
\emph{Behav. Res. Methods}, vol. 53, no. 4, pp.~1689--1696, Aug. 2021, doi: 10.3758/s13428-020-01516-y.

\bibitem{bib61}\label{bib61} M. {\'A}. Goda and P. H. Charlton and J. A. Behar, ``pyPPG: a Python toolbox for comprehensive photoplethysmography signal analysis,'' {\it Physiological Measurement}, vol. 45, no. 4, pp. 045001, Apr. 2024, {{\it IOP Publishing}} doi: 10.1088/1361-6579/ad33a2.

\bibitem{bib25}\label{bib25} H. Cao, Y. Wang, J. Chen, D. Jiang, X. Zhang, Q. Tian, and M. Wang, 
``Swin-Unet: Unet-Like Pure Transformer for Medical Image Segmentation,'' 
in \emph{Computer Vision – ECCV 2022 Workshops}, Tel Aviv, Israel, Oct. 2022, 
Lecture Notes in Computer Science, vol. 13841, Springer, 2023, pp.~205--218, 
doi: 10.1007/978-3-031-25066-8\_9.

\bibitem{bib30}\label{bib30} V. Sanh and L. Debut and J. Chaumond and T. Wolf, ``DistilBERT, a distilled version of BERT: smaller, faster, cheaper and lighter,'' arXiv, 2020

\bibitem{bib60}\label{bib60} O. Ronneberger, P. Fischer, and T. Brox, 
``U-Net: Convolutional Networks for Biomedical Image Segmentation,'' 
in \emph{Medical Image Computing and Computer-Assisted Intervention -- MICCAI 2015}, 
Lecture Notes in Computer Science, vol. 9351, Cham: Springer, 2015, pp.~234--241, 
doi: 10.1007/978-3-319-24574-4\_28.
\end{thebibliography}
\end{document}